\documentclass[preprint]{elsarticle}
\usepackage{amssymb}
\biboptions{sort&compress}
\usepackage{comment}
\usepackage{graphicx} 
\usepackage{subcaption}
\usepackage{gensymb}
\usepackage{color}
\usepackage{amsmath}
\usepackage{float}
\usepackage{soul}
\usepackage{longtable}
\usepackage{multirow}
\usepackage{hyperref} 
\usepackage{lmodern,blindtext}

\journal{Thermal Science and Engineering Progress}
\begin{document}
\begin{frontmatter}


\title{A one-dimensional numerical model for system performance prediction of loop heat pipes and its validation \tnoteref{published}}
\tnotetext[published]{The final version of this paper can be found in \textcolor{red}{https://doi.org/10.1016/j.tsep.2025.103501}.}

\author[inst1]{Xu Huang}\corref{cor1}
\ead{xu.huang@kuleuven.be}

\author[inst1,inst2,inst3]{Geert Buckinx}
\author[inst1,inst3]{Maria Rosaria Vetrano}
\author[inst1,inst3]{Martine Baelmans}

\affiliation[inst1]{organization={Department of Mechanical Engineering, KU Leuven},
            addressline={Celestijnenlaan 300A}, 
            city={3001 Leuven},
            country={Belgium}}
            
\affiliation[inst2]{organization={Vito},
            addressline={Boeretang 200}, 
            city={2400 Mol},
            country={Belgium}}

\affiliation[inst3]{organization={EnergyVille},
            addressline={Thor Park}, 
            city={3600 Genk},
            country={Belgium}}
            
\cortext[cor1]{Corresponding author}

\begin{abstract}
Developing high power density electronics requires effective and highly reliable cooling techniques with low thermal resistance and high heat removal capacity. Loop heat pipes (LHPs) are one kind of two-phase heat transfer device which can meet all these requirements. 
A physics-based one-dimensional numerical model has been elaborated to further develop this promising electronics cooling solution. In addition, an experimental setup using water as coolant, and which includes an LHP built with transparent materials, is used to validate the numerical model. This validation is obtained by comparing the numerical results with the temperature measurements, the two-phase flow section length in the condenser line, and an energy balance evaluation. The numerical model is then used to predict the system performance of the LHP under investigation. The results indicate that this LHP can only work in fixed conductance mode for the given operating conditions. Further, the influence of the charging mass is assessed. For low charging masses, both the natural convection cooled condenser and the compensation chamber casing can be fully utilized to dissipate the input heat, which can exceed {29.9 W} when the coolant saturation temperature is at {100 \celsius}. 
The operational limits for LHPs are investigated. For the current LHP under given operating conditions, the heat dissipation limit restricts the maximum input heat power, which in turn induces the activation of the heat leakage limit and the liquid-filling limit. It is shown that the condenser with higher heat dissipation capacity and lower thermal resistance will therefore enlarge the operational envelope.
\end{abstract}


\end{frontmatter}

\section{Introduction}
\label{sec: introduction}
The continuous miniaturization of electronic devices comes with a need for more effective and reliable cooling methods. The main reason is that the generated heat per chip area vastly increases as the number of transistors per unit chip area continues to grow. This leads to higher cooling rate requirements while keeping the die temperature as well as thermal stresses induced by non-uniform temperature distributions in the chip within their technical limits \cite{schelling2005managing}. To mitigate these problems, novel cooling methods based on two-phase heat transfer are becoming more and more attractive. Especially, loop heat pipes (LHPs) are among the most promising ones \cite{maydanik2005miniature}.

As a response to the demanding needs in thermal control of spacecrafts, the LHP has been invented and gradually became a standard component in space missions \cite{riehl2005development}. 
Compared to conventional heat pipes, which have been widely embedded in solutions for electronics cooling, LHPs not only have a higher cooling rate but can also transport heat over longer distances. This is accomplished by capillarity forces that are realized in a dedicated wick structure. Moreover, their flexible pipe structure makes them very suitable for deployment in devices with limited space \cite{maydanik2005miniature,maydanik2005loop}. 

\begin{figure}[h] 
\centering
\includegraphics[width=0.78\textwidth, angle = 0]{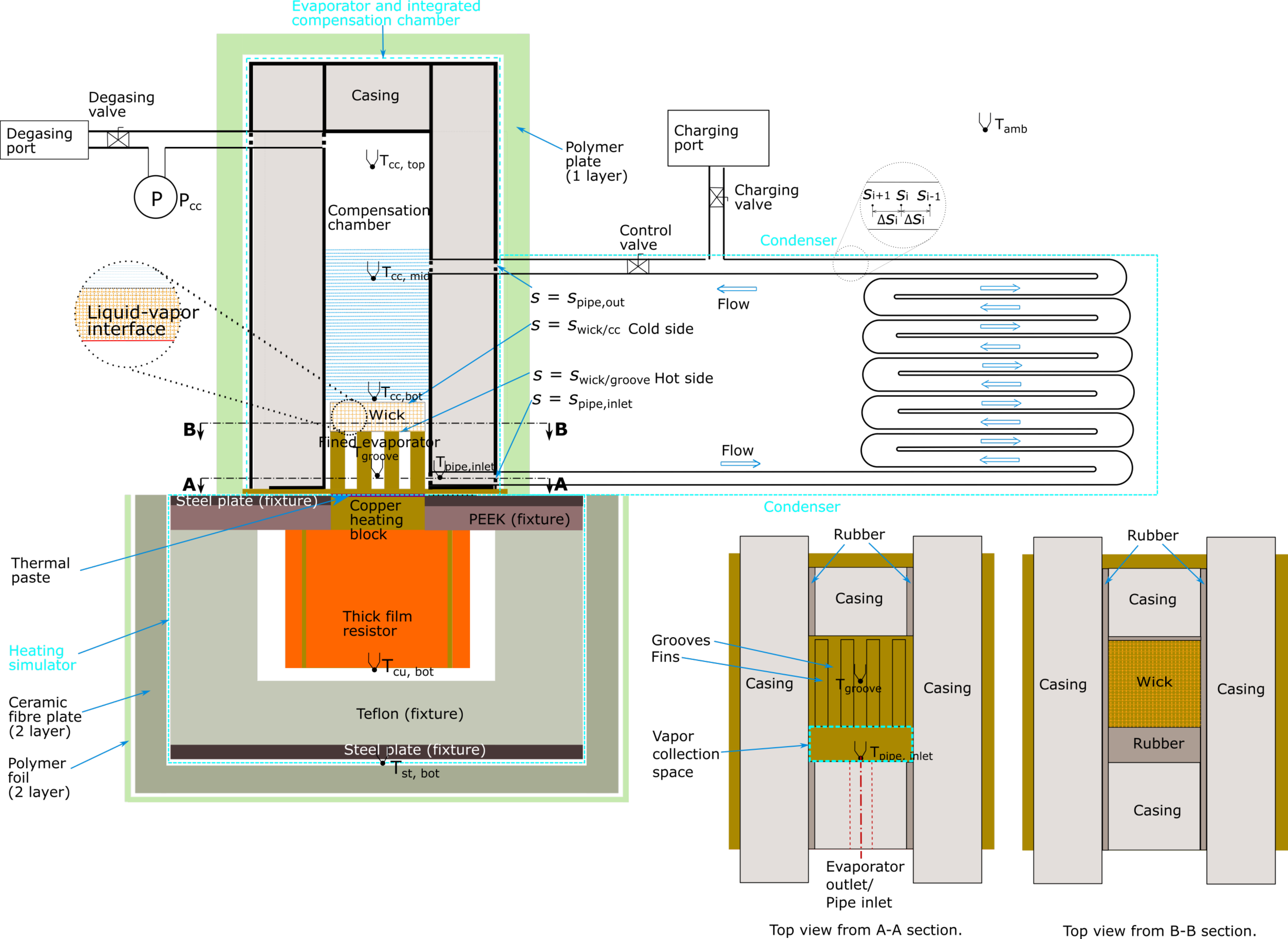}
\caption{Illustration diagram of the experimental setup: the LHP system integrated with the heating simulator, the embedded thermocouples and pressure transducers.}
\label{fig01}
\end{figure}

A typical LHP consists of an evaporator, a vapour line, a condenser line, and a liquid line. The evaporator includes a porous wick structure and a compensation chamber. The system is shown in Fig. \ref{fig01}. 
Usually, the evaporator has a finned surface to conduct the heat from the heat source toward the coolant. 
As the coolant takes up the heat, it evaporates in the space between the evaporator fins. 
The evaporating coolant remains separated from the single-phase liquid coolant through the porous wick. 
This porous wick generates a capillary force, which drives the coolant circulation and overcomes the pressure losses in the LHP \cite{maydanik2005loop}. 
The vapour is brought back to the compensation chamber after being condensed in the condenser. The compensation chamber acts as a liquid reservoir, storing excess coolant during operation and supplying liquid coolant to the wick. 
Two typical operating modes have been reported for the established LHPs, namely the variable conductance mode (VCM) and the fixed conductance mode (FCM) \cite{maydanik2005loop,ku1999operating}.
The VCM mode of the LHPs is associated with a varying condensation length in the condenser with the input power. The FCM is associated with a constant condensation length and is featured by a quasi-linearly increase in the operating temperature
with the increased input power.

Due to the advantages over other heat transfer devices, the LHPs have been widely investigated to extend their applicability for electronics cooling by miniaturizing the evaporator \cite{maydanik2005miniature,maydanik2020investigation,ramasamy2018miniature}. Most designs make use of a cylindrical evaporator, with diameter that can be small as {6 mm}. Using forced convective cooling outside their condenser, the well-designed system can even transport heat of {100-200 W} \cite{maydanik2005miniature}. The thorough experimental tests further prove the LHPs' ability in fitting various operating conditions in electronics cooling \cite{maydanik2020investigation,ramasamy2018miniature,zilio2018active,zhang2023performance,zhao2023r1234ze}.

The use of cylindrical evaporators has the advantage of good mechanical strength, which is crucial due to the potential high pressure inside the loop, especially when using ammonia as a coolant.
However, a squared or flat evaporator is preferred because of its more easy integration in electronic devices. Moreover, when using cylindrical evaporators, a special thermal interface or so-called `saddle' is mandatory \cite{maydanik2005loop}, as shown in \cite{maydanik2005miniature,maydanik2020investigation,ramasamy2018miniature,zilio2018active,maydanik2022visual}. This extra saddle limits miniaturization and poses extra thermal resistance between the heat-generating surface and the evaporator. In addition, with its varying thickness to fit the shape of the cylindrical evaporator its thermal resistance is unevenly distributed. This may lead to an uneven temperature distribution on the devices' surface, further leading to thermal stresses. In contrast, the flat evaporator can be directly placed on the heat-generating surface of devices. Therefore, in this paper we focus on this configuration.

Singh et al. \cite{singh2007miniature} developed a LHP with a flat (disk-shaped) miniature evaporator having a surface area of {7.1 ${\rm cm^2}$}. Using forced cooling at the condenser exterior, the system can transport heat load up to {70 W}. 
Li et al. \cite{li2010experimental} developed an LHP with a flat evaporator and forced air cooling for the condenser. Tested vertically with the condenser above the evaporator, this LHP can handle a heat flux up to {100 ${\rm W/cm^2}$}.
Chernysheva et al. \cite{chernysheva2014copper} built a copper-water LHP with a flat evaporator that can be applied to cooling systems of supercomputers. The LHP's operation characteristics with different cooling water temperatures were investigated using forced water cooling in the condenser. 
Tian et al. \cite{tian2019experimental} also developed one LHP with a flat disk type miniature evaporator, which can be started up successfully under a heat load of {2 W} and works under a heat load of up to {60 W}. The setup used forced water cooling to remove heat from the condenser.
Zhao et al.\cite{zhao2023r1234ze} developed two LHPs with flat disk evaporator for electronics cooling. The coolant \textit{R1234ze(E)} was chosen due to, amongst other, its advantageous low saturation temperature. Both air and liquid were used in the forced cooling outside the condensers, so that the LHP system was able to transport up to {$270$ W} of heat power without deprime.
Some LHPs with ultra-thin evaporators were also presented \cite{fukushima2017new, zhou2019power,phan2022novel}. They are designed for laptops and can typically transport a maximum heat load of around {10 W}. In this work we study a square evaporator/wick structure with a surface of {1 ${\rm cm^2}$}, while passively cooling the LHP by natural convection and radiation.

The design of LHPs can be facilitated by employing numerical models, which can reduce the design cost and time compared to an experimental approach. Different one-dimensional models have been presented to predict the system performance of LHPs.
Chuang et al. \cite{chuang2002comparison} presented a one-dimensional model for an LHP, which was then compared with their own experimental data. 
The model ignores the heat dissipation from the condenser through thermal radiation and assumes that all heat from the heater was put into the setup. 
Because the author used ammonia as a coolant, the operating temperature was around {30 \celsius} which legitimates these assumptions. 
The model successfully reproduced the LHP performance when the setup operates in an horizontal position or under adverse elevation (condenser below the evaporator), leading to a discrepancy as low as {2\%}. However, it failed to predict the temperature in case of positive elevation (condenser positioned above the evaporator).
Launay et al. \cite{launay2008analytical} presented an analytical expression to simulate the LHP performance aiming at a sensitivity analysis of various parameters.
Adoni et al. \cite{adoni2009effects} established a theoretical model to predict the steady state performance of an LHP. Both the model and the experimental results suggested that the system works in a fixed conductance mode since the large liquid inventory in the compensation chamber resulted in the loss of thermal link between the cylindrical evaporator core and the compensation chamber, resulting in a constant condenser length usage. 
Bernagozzi et al. \cite{bernagozzi2018lumped} built a one-dimensional lumped parameter model to simulate the behaviour of LHP to be applied to electric vehicles. The model was also used to comprehensively study the influence of various geometrical parameters on system performance. This unveiled the potential of applying LHPs in electric vehicles. 
Li et al. \cite{li2022steady} used a one-dimensional model to check the steady state performance of LHPs with flat evaporator. The influence of the gravity was included in the model to check the operating characters of the system when condenser was arranged above the evaporator. Both the VCM and the FCM were recognized in the results.

In the literature, almost all the available LHP models have been elaborated for forced convection outside the LHPs' condenser, as are the most common in-house setups. 
Numerical models for setups using natural convection at the outside of the condenser are relatively rare, even though a passively cooled LHP is very attractive, even when the corresponding  cooling is typically lower than their forced cooled counterparts.
Consequently, the model validation will be very sensitive to various parameters, like the input power, heat transfer coefficient, etc. 
As these parameters all depend on the measured temperature, highly accurate temperature measurements are in turn necessary. Traditionally, temperature measurements are mostly conducted at the outer surface of the LHP. 
This provides temperatures that differ from the coolant's actual temperature, and may hamper a correct validation when natural convection is used. 
In addition, the  numerical models in the literature are frequently based on the assumption that the coolant in the compensation chamber is either only in a liquid phase \cite{ramasamy2018miniature}, or saturated (i.e. both liquid and vapor phase are present simultaneously in the compensation chamber) \cite{fukushima2017new,launay2008analytical,bernagozzi2018lumped,meinicke2019lean}. However, it has been reported that the coolant in the compensation chamber can switch between two situations during the LHP operation\cite{chernysheva2014copper,launay2008analytical,adoni2007thermohydraulic,chen2021heat}. This also limits the applicability of the previous models.

In this paper, we first develop a 1D numerical model for a passively cooled LHP system. Therefore, in the next section, the key equations and assumptions used in the model are provided. The solution procedure for the model and the operational limits are explained in section \ref{sec: solution}.  
Using this model, we design and build an experimental setup that will serve to validate the assumptions and the results of the obtained numerical simulations. Section \ref{sec: experiment setup and results} gives details about the experimental setup, experimental results, and data post-processing. Section \ref{sec: results and discussion} focuses on the parameter calibration and model validation procedure. It is followed by a discussion on the numerical results of the model, including performance curves and operational limits of the current setup.

\section{LHP model} 
\label{sec: lhp model} 
\subsection{Main modelling assumptions}
\label{subsec: main modelling assumptions}
To establish our model, the following assumptions have been made.
First, we assume that all state variables like temperature, pressure, and flow velocity vary primarily in the main flow direction allowing us to assess the LHP performance with a 1D model.
The coolant flow has steady state conditions that are reached instantaneously for every thermal operation condition. 
The compressibility of the vapor phase in the two-phase flow section is neglected \cite{carey2020liquid}, as the pressure variation along the condenser is low.
We assume the wick to be completely filled with liquid coolant so that the liquid-vapor interface coincides with the wick-groove boundary. Consequently, we model the coolant inside the grooves as fully saturated homogeneous vapor and neglect heat convection and pressure drops due to the coolant flowing inside the evaporator grooves.
In addition, we model the entire pipe line between the evaporator and the compensation chamber as one component with three regions representing the vapor line, condenser, and liquid line.
Further, the heat losses towards the ambient from the outer surface of the evaporator's casing, and thus from the wick and compensation chamber, are lumped together as a single total heat loss.
The main part of the thermal inertia in our system is linked with the heat source and the evaporator (around {88\%}). Therefore, to obtain the \textit{effective} input power $\dot{Q}_{\text{eff}}$ to the coolant cycle, we take this thermal energy stored in the whole heater and the LHP system into account and subtract it from the total input power. 
Finally, we ignore the influence of gravity, as the LHP is put in a horizontal position.

\subsection{Evaporator model} 
\label{subsec: evaporator model}
In the following, we focus first on modeling the evaporator's core element, i.e., the wick structure. Subsequently, we bring in the necessary boundary conditions and additional equations. 

In the evaporator wick, the temperature $T_l$ of the liquid coolant varies with the main flow direction $s$, and is governed by the energy equation
\begin{equation} \label{eq: thermal equation coolant in wick}
\dot{m} c_{P,l} \frac{dT_l}{ds} = k_{\text{eff},l}  A_{\text{wick}} \frac{d^2T_l}{ds^2}
\end{equation}
Here, $\dot{m}=\rho_lu_lA_{\text{wick}}$ is the mass flow rate through the system, with $\rho_l$, $c_{P,l}$, and $u_l$ denote the coolant's density, specific heat capacity, and flow velocity in the liquid phase ($l$), and $A_{\text{wick}}$ is the wick cross sectional area.
As $T_l$ represents the homogenized or macro-scale temperature of the liquid coolant in the porous wick, an effective thermal conductivity $k_{\text{eff},l}$ appears in equation (\ref{eq: thermal equation coolant in wick}).

At the liquid-vapor interface along the wick-groove boundary, the temperature of the liquid coolant equals the saturation temperature of the vapor in the grooves. Hence the operating temperature $T_{\text{op}}$, as is defined in \cite{chernysheva2007operating}, is finally determined by the vapor pressure $P_{v}$ in the grooves 
\cite{figus1999heat,kaya2006numerical}:
\begin{equation}
T_l=  T_{sat} (P_{v})  = T_{\text{op}}
\qquad \mbox{at $s=s_{\text{wick/groove}}$}
\end{equation}
Furthermore, due to energy conservation across the vapor-liquid interface, the gradient of the liquid temperature satisfies
\begin{equation}
\label{eq: energy balance at wick/groove boundary}
k_{\text{eff},l} A_{\text{wick}} \frac{dT_l}{ds}  = - \dot{m}H_{\text{lat}}  + \dot{Q}_{\text{eff}}
\qquad \mbox{at $s=s_{\text{wick/groove}}$} \,,
\end{equation}
where $H_{\text{lat}}$ is the specific latent heat, which depends on the coolant temperature $T_{\text{op}}$ at the wick-groove boundary.
By using $\dot{Q}_{\text{eff}}$ at the wick-groove boundary, we ignore the thermal resistance of the evaporator fins and the heat convection between the fins and imagine that the hot side of the wick is directly in contact with the heat source.

At the wick cold side, the gradient of the liquid temperature in the wick contributes to the energy balance for the fluid in the compensation chamber (cc):
\begin{equation}
\label{eq: energy balance cc}
k_{\text{eff},l} A_{\text{wick}} \frac{dT_l}{ds} = \dot{m}  \left(
H_{\text{wick/cc}} - H_{\text{cond,outlet}}
\right) + 
\dot{Q}_{\text{cc}} 
\qquad \mbox{at $s=s_{\text{wick/cc}}$}
\end{equation}
Here, $H_{\text{wick/cc}}$ refers to the specific enthalpy $H= c_{P,l}T_l$  of the coolant leaving the compensation chamber as sub-cooled or saturated liquid, which is evaluated based on the liquid temperature $T_l$ at $s=s_{\text{wick/cc}}$. 
Similarly, $H_{\text{cond,outlet}}$ denotes the specific enthalpy of the single- or two-phase coolant entering the compensation chamber at $s=s_{\text{cond,outlet}}$.
The mass flow rate $\dot{m}$ through the compensation chamber equals the mass flow rate through the entire LHP, as we assume quasi-steady operation. 

The heat loss from the compensation chamber towards the ambient, $\dot{Q}_{\text{cc}}$, in equation (\ref{eq: energy balance cc}), 
follows from the mean temperature difference between the compensation chamber and the ambient:
\begin{equation} \label{eq: heat loss cc}
\dot{Q}_{\text{cc}} = \frac{T_{\text{cc}} - T_{\text{amb}} }{ R_{\text{cc}}}  
\end{equation}
The temperature in the compensation chamber is supposed to be uniform so that $T_{\text{cc}}$ is equal to the liquid temperature $T_l$ at $s=s_{\text{wick/cc}}$.
The thermal resistance $R_{\text{cc}}$ has been determined assuming that the heat loss occurs via perpendicular conduction through the top and sides of the evaporator casing and the surrounding insulation material.
In addition, we assume that natural convection at the outer surface of the insulation material takes place, which we model using the heat transfer coefficients for vertical and horizontal walls from \cite{lienhard2005heat,goswami2004crc}, as explained in \ref{app: thermal resistance from coolant in cc to ambient}.

The pressure gradient in the wick is governed by $\dot{m}$, via the wick's permeability $K$, according to Darcy’s law \cite{chuang2002comparison}:
\begin{equation} 
\label{eq: momentum equation evaporator}
\frac{dP_l}{ds} = - \frac{u_l \mu_l}{K}  = - \frac{\dot{m} \mu_l}{\rho_l K A_{\text{wick}}}
\end{equation}
The previous differential equation governs the pressure $P_{l}$ of the liquid coolant in the wick, which satisfies the boundary condition
\begin{equation}
P_l = P_{\text{cond,outlet}}
\qquad
\mbox{at $s=s_{\text{wick/cc}}$}\,,
\end{equation}
since the pressure in the compensation chamber is considered to be uniform and equal to the pressure at the condenser outlet, $P_{\text{cond,outlet}}$.

The model equations presented above are complemented with the thermodynamic state functions from {\rmfamily\scshape Refprop} \cite{lemmon2007nist}, which relate the density $\rho$, enthalpy $H$, and other coolant properties like $H_{\text{lat}}$ to the temperature and pressure at each position in the evaporator.
That way, they can be solved to obtain the pressures and temperatures $P_l(s)$ and $T_l(s)$ along the main flow direction in the wick and compensation chamber, as well as the mass flow rate $\dot{m}$, if the operating temperature $T_{\text{op}}$, the effective input power $\dot{Q}_{\text{eff}}$, and the inlet conditions for the evaporator ($P_{\text{cond,outlet}}$, $H_{\text{cond,outlet}}$) are specified.

\subsection{Condenser model}
\label{subsec: condenser model} 
The coolant flow through the condenser, which includes vapor and liquid lines, is divided into a single-phase vapor region ($v$), a single-phase liquid region ($l$), and a two-phase region ($lv$) in between. 
In all regions the coolant enthalpy $H_i$ satisfies the energy balance
\begin{equation}
\label{eq: energy balance single-phase regions condenser}
\dot{m}  \frac{dH_i}{ds}  = - \dot{Q}^{'}_{\text{cond}} \,, 
\end{equation}
where the specific enthalpy $H_{lv}$ of the two-phase coolant is related to that of the pure vapor and liquid phases via the vapor quality $x$, as $H_{lv} = (1-x) H_{l}(T_{\text{sat}}) + x H_{v}(T_{\text{sat}})$, with the local saturation temperature $T_{lv}=T_{\text{sat}}(P_{lv})$ evaluated at the local pressure $P_{lv}$. 
Further, $\dot{Q}^{'}_{\text{cond}}$ denotes the heat per unit of condenser length, which is transferred from the coolant towards the ambient at each condenser section with coordinate $s$. It can be written as 
\begin{equation}
\dot{Q}^{'}_{\text{cond}} = \frac{T_{i} - T_{\text{out}}}{R^{'}_{\text{cond,in}}} \,,
\end{equation}
where 
$R^{'}_{\text{cond,in}} = {1}/{(h_{\text{in}}  \pi D_{\text{in}})} 
+ 
{[ \ln \left(D_{\text{out}}/ D_{\text{in}} \right) ]}/{( 2\pi k_{\text{cond}} )}$ 
is the thermal resistance per unit length between the coolant at temperature $T_i$ and the condenser's outer surface at temperature $T_{\text{out}}$.
It is composed of the single- and two-phase heat convection coefficient $h_{\text{in}}$ in the condenser pipe, and the thermal conductivity $k_{\text{cond}}$ of the condenser walls. 

At the outer surface of the condenser, the same heat transfer rate is the sum of two contributions: 
$\dot{Q}^{'}_{\text{cond}} = \dot{Q}^{'}_{\text{cond,conv}} + 
\dot{Q}^{'}_{\text{cond,rad}}$.
The first contribution, 
\begin{equation}\dot{Q}^{'}_{\text{cond,conv}} = \frac{T_{\text{out}} - T_{\text{amb}}}{R^{'}_{\text{cond,conv}}} \,,
\end{equation}
is due to the natural heat convection at the outer surface of the condenser pipe.
It is characterized by the heat convection coefficient $h_{\text{out}}$ and the thermal resistance per unit length $R^{'}_{\text{cond,conv}} = {1}/{( h_{\text{out}}  \pi D_{\text{out}} )}$.

The convective heat transfer coefficient inside the condenser, $h_{\text{in}}$, is modelled under the assumption of a constant heat flux at the condenser wall, using \textit{Nusselt}'s correlation $h_{\text{in}} = 3.66 \Phi k_i/D_{\text{in}}$ for the single-phase laminar regime ({$Re_i <$ 2100}), and the modified \textit{Dittus-Boelter} correlation for the single-phase turbulent regime ({$Re_i >$ 2100}), as well as the two-phase region: $h_{\text{in}} = 0.023 \xi_{\text{in}} \Phi Re_i^{0.8} Pr_i^{0.3}$ $k_i / D_{\text{in}}$ \cite{winterton1998did}.
The latter is based on the \textit{Reynolds} number $Re_i= \rho_i u_i D_{\text{in}}/\mu_i$ and the \textit{Prandtl} number $Pr_i$ of the local phase $i$, as well as an empirical tuning factor $\xi_{\text{in}}$, which will be explained further.
The two-phase multiplier $\Phi$ equals $1$ in the single-phase regions ($i=\{v,l\}$).
In the two-phase region ($i=lv$), it follows from the local vapor quality $x$ and the reduced pressure $P_r$ at each section, according to \textit{Shah}’s correlation \cite{shah1979general}: $ \Phi =(1-x)^{0.8} + 3.8 x^{0.76} (1-x)^{0.04}P_r^{-0.38}Pr^{0.1}$, if one takes the liquid phase $i=l$ as reference. 

The convective heat transfer coefficient at the outer surface of the condenser, $h_{\text{out}}$, is modelled based on \textit{Morgan}’s correlation \cite{morgan1975overall} for natural heat convection by air over the outer surface of a horizontal cylindrical pipe: $h_{\text{out}} = \xi_{\text{out}} C Ra^n k_{\text{air}}/ D_{\text{out}}$.
In this correlation, the \textit{Rayleigh} number $Ra$ for the air flow is based on the diameter $D_{\text{out}}$ and the local temperature difference $T_{\text{out}} - T_{\text{amb}}$ at the specified position $s$. An empirical tuning factor $\xi_{\text{out}}$ is also involved and will be explained further.
The constants $C$ and $n$ are given by {$C$ = 0.675} and {$n$ = 0.058} when {${\rm 10^{-10}}$ $< Ra <$ ${\rm 10^{-2}}$}, while {$C$ = 1.02} and {$n$ = 0.148} when {${\rm 10^{-2}}$ $< Ra <$ ${\rm 10^{2}}$}.
For {${\rm 10^{2}}$ $< Ra <$ $ {\rm 10^{4}}$}, we have {$C$ = 0.850} and {$n$ = 0.188} \cite{morgan1975overall}.

The second contribution to $\dot{Q}^ {'}_{\text{cond}}$ is due to the radiative heat transfer from the condenser's outer surface towards the ambient, which is modelled using a lumped thermal resistance \cite{lienhard2005heat}:
\begin{equation}
\label{eq: condenser radiation}
\dot{Q}^{'}_{\text{cond,rad}}=  
\left(
\frac{1-\epsilon}{\epsilon} + 
\frac{1}{F}
\right)^{-1}
\pi D_{\text{out}} \sigma_{\text{SB}} 
\left(T_{\text{out}}^4- T_{\text{amb}}^4\right)
\end{equation}
The radiative heat transfer is described by the \textit{Stefan-Boltzmann} constant $\sigma_{\text{SB}}$, the emissivity of the condenser's cylindrical surface $\epsilon$, and its view factor $F$ with respect to the ambient.
The ambient itself is regarded as a black body here.

To determine the thermodynamic state in the condenser,  the pressure distribution $P_i$ is evaluated using the momentum equation at every section $s$:
\begin{equation}
\label{eq: momentum equation condenser}
\frac{dP_i}{ds} = -  \frac{\pi D_{\text{in}}}{A_{\text{in}}}  \tau_{w,i} - \frac{d P_{\text{acc}}}{ds}
\quad
\mbox{with} 
\quad
\tau_{w,i} = \Phi_{P,i}^2  \frac{f_i}{8} \rho_i u_i^2
\end{equation}
In the single-phase regions $i=\{l,v\}$, where $\Phi_{P,i}=1$ and $d P_{\text{acc}}/ds=0$, the pressure $P_i$ at section $s$ is thus obtained from the mass flow rate $\dot{m}= \rho_i u_i A_{\text{in}}$ and the corresponding \textit{Darcy} friction factor $f_i$. 
The latter is given by the \textit{Hagen–Poiseuille} equation $f_{i} = 64/Re_{i}$ for laminar flow ({$Re_{i} <$ 4000}), and the \textit{Blasius} correlation $f_{i} = 0.316/Re_{i}^{0.25}$ for turbulent flow ({$Re_{i} >$ 4000}).
In the two-phase region where $i=lv$, 
the velocity $u_{i}$ and \textit{Reynolds} number $Re_{i}$ for each phase are evaluated assuming that the corresponding phase flows through the entire cross-sectional area of the condenser.
The pressure $P_{lv}$ is then obtained from the two-phase multiplier $\Phi_{P,i}$ and the acceleration pressure drop $d P_{\text{acc}}/ds$.

The two-phase multiplier $\Phi_{P,i}$ is calculated from the \textit{Lockhart-Martinelli} correlations \cite{lockhart1949proposed,chisholm1967theoretical} based on the \textit{Martinelli} parameter $X = [(d P_{l}/ds)/(d P_{v}/ds)]^{0.5}$ as in \cite{chuang2002comparison}, where $P_{l}$ and $P_{v}$ represents the friction pressure drop assuming the liquid and vapor flow through the entire cross-section of the condenser with their respective mass flow rates.
So, depending on whether one takes the density $\rho_i$ and flow velocity $u_i$ of the liquid phase ($i=l$) or vapour phase ($i=v$) as reference in the expression for $\tau_{w,lv}$, the latter follows from
\begin{align}
\Phi_{P,l}^2 &= 1 + \frac{C_{\text{Ch}}}{X} + \frac{1}{X^2}
\qquad \mbox{for $i = l$ when $Re_l > 4000$} \,,\\
\Phi_{P,v}^2 &= 1 + C_{\text{Ch}}X + X^2 
\qquad \mbox{for $i = v$ when $Re_l < 4000$} \,,
\end{align}
where the \textit{Chisholm} parameter $C_{\text{Ch}}$ is a constant which depends on the flow regime of each phase \cite{chisholm1967theoretical}.
The acceleration pressure drop, on the other hand, follows from the mass flux of coolant $G = \dot{m}/A_{\text{in}}$ in the condenser and the local void fraction $\alpha$ at each section:
\begin{equation}
\frac{d P_{\text{acc}}}{ds} = 
G^2 
\frac{d}{ds} 
\left[
\frac{(1-x)^2}{\rho_l(1- \alpha)} + 
\frac{x^2}{\rho_v \alpha} 
\right]
\end{equation}

The void fraction has been linked to the \textit{Martinelli} parameter $X$ using \textit{Butterworth}'s correlation $\alpha = (1 + 0.28 X^{0.71})^{-1}$ \cite{lockhart1949proposed,butterworth1975comparison}. It should be noted that for full consistency with the energy equations also the difference in kinetic energy should be retained in equation (\ref{eq: energy balance single-phase regions condenser}). However this term is negligible in the energy balance compared to the condenser cooling rate. On the other hand, it could be worthwhile to consider the incorporation of the acceleration pressure drop in the evaporator model too. 

The differential equations (\ref{eq: energy balance single-phase regions condenser}) and (\ref{eq: momentum equation condenser}) can be solved together with the thermodynamic state functions from {\rmfamily\scshape Refprop} to obtain the complete temperature and pressure distributions in the condenser, for the given inlet condition $T_{v} = T_{\text{sat}}$ at $s=s_{\text{cond,inlet}}$.

It is clear that overall balances are easily retrieved based on the discretized equations. As such, the overall energy balance for the condenser reads, 
\begin{equation} 
\label{eq: energy balance condenser}
\dot{Q}_{\text{cond}} = \int_{s_{\text{cond,inlet}}} ^{s_{\text{cond,outlet}}} \dot{Q}^{'}_{\text{cond}}{ds}
=
\dot{m} 
\left(H_{\text{cond,inlet}} - H_{\text{cond,outlet}}
\right) \,,
\end{equation}
where $H_{\text{cond,inlet}}(T_{\text{sat}})$ and $H_{\text{cond,outlet}}$ are the coolant's specific enthalpy at the condenser inlet and outlet respectively.
Furthermore, when the previous condenser and evaporator model are solved together as a system of equations for the entire LHP, the temperature and pressure distributions will automatically respect the following energy balance for the entire LHP:
\begin{equation}
\label{eq: energy balance LHP}
\dot{Q}_{\text{eff}} = 
\dot{Q}_{\text{cond}} + \dot{Q}_{\text{cc}}
\end{equation}

\subsection{Charging mass of coolant}
\label{subsec: charging mass} 
From the preceding model equations, the total charging mass of coolant $M_{\text{fill}}$ in the LHP can be obtained as the sum of the coolant masses in each subsystem:
\begin{equation}
\label{eq: total mass in system}
M_{\text{fill}}  =
M_{\text{groove}} + 
M_{\text{wick}} +
M_{\text{cc}} + 
M_{\text{cond}}
\end{equation}

The coolant mass in the evaporator groove volume $V_{\text{groove}}$ (i.e. the space between the fins and the accumulator) is simply given by $M_{\text{groove}} = \rho_v V_{\text{groove}}$, with $\rho_v=\rho_v(T_{\text{op}})$ the density of the saturated vapor.

The mass of coolant inside the wick follows from the porosity $\phi$ and the total volume $V_{\text{wick}}$ of the wick:
\begin{equation}\label{eq: mass in wick}
M_{\text{wick}} =  \phi A_{\text{wick}} \int_{s_{\text{wick/cc}}} ^{s_{\text{wick/groove}}} \rho_{l}  ds
\simeq  \phi \rho_{l} V_{\text{wick}} \,,
\end{equation}
since the liquid density $\rho_{l}$ remains virtually constant in the wick.

The mass of coolant in the condenser is evaluated by integration of the void fraction and density along the condenser: 
\begin{equation}\label{eq: mass in pipe}
M_{\text{cond}} = A_{\text{in}} \int_{s_{\text{cond,inlet}}} ^{s_{\text{cond,outlet}}} \left[ \rho_{v}  \alpha + \rho_{l}  (1 - \alpha) \right]  ds
\end{equation}

Finally, the coolant mass stored in the compensation chamber with volume $V_{\text{cc}}$ equals 
\begin{equation} 
\label{eq: mass in cc}
M_{\text{cc}} = \alpha_{\text{cc}}(V_{\text{cc}} - V_{\text{NCG}})\rho_v + (1- \alpha_{\text{cc}})(V_{\text{cc}} - V_{\text{NCG}})\rho_l
\end{equation}
Here, $V_{\text{NCG}}$ corresponds to the volume occupied by non-condensable gasses, which accumulate in the compensation chamber, as they hardly dissolve in the coolant fluid and barely mix with the coolant vapor. 
The latter has to be supplied to the model in order to agree with the experiment (see section \ref{subsubsec: model validation and discrepancy}).
We note that 
$\alpha_{\text{cc}}$, $\rho_v$ and $\rho_l$ are uniform in the compensation chamber.

The value of $\alpha_{\text{cc}}$ is implicitly related to the state of coolant in the compensation chamber \cite{chernysheva2007operating,adoni2007thermohydraulic}:
when a saturated mixture of vapor-liquid coolant is present, $\alpha_{\text{cc}} > 0$ must hold, as $T_{\text{cc}} = T_{\text{sat}}(P_{\text{cc}})$ , or equally $P_{\text{cc}} = P_{\text{sat}}(T_{\text{cc}})$. 
When the compensation chamber is completely filled
with either saturated or subcooled liquid,
$\alpha_{\text{cc}} = 0$ must hold, as $T_{\text{cc}} \leq T_{\text{sat}}(P_{\text{cc}})$.

\subsection{Operational limits of the LHP}
\label{subsec: limits} 
There are nine operational limits embedded in the model: the viscous limit, the sonic limit, the entrainment limit, the capillary limit, the liquid-filling limit, the heat-leakage limit, the heat-dissipation limit, the coolant limit, and finally, the application temperature limit. 
For the definition of the first four limits, we refer to the literature \cite{launay2007parametric,chernysheva2008heat}.

In our model, we have incorporated the entrainment limit using the 
analytical expressions from \cite{chi1976heat,riffat2002analytical}.
For the sonic limit, we use the analytical expression from \cite{chi1976heat,ZhuangJun2000bookHeatPipe}, in which we use the maximum pore size measured at the upper surface of the wick. Both limits are checked by comparing the effective input power $\dot{Q}_{\text{eff}}$ with the maximum allowed input power evaluated from these analytical expressions.
Although it is common practice to define the viscous limit as the requirement that the condenser inlet pressure should exceed the coolant's triple point pressure ($P_{\text{cond,inlet}} > P_{\text{triple}}$), we adopted the criterion that the absolute pressure $P_i$ at every point in the LHP should exceed $P_{\text{triple}}$.

The capillary limit is incorporated in our model by checking whether the pressure drop along the entire loop is lower than the maximum capillary pressure, i.e., $P_v - P_l = \Delta P_{\text{cap}} \leq \Delta P_{\text{cap,max}}$ at $s=s_{\text{wick/groove}}$.
The latter is calculated from the maximum pore size at the upper surface of the wick $r_{\text{pore}}$, the interfacial surface tension $\sigma$
and the contact angle in the wick pores $\theta = 52\degree$, according to 
$\Delta P_{\text{cap,max}} = 2\sigma\cos\theta / r_{\text{pore}}$ \cite{chernysheva2008heat}.

The liquid-filling limit is identified by checking that the void fraction should remain below $0.5$ in the compensation chamber, and above zero in all other components: $\alpha_{\text{cc}} < 0.5$ and  $\alpha > 0$.
Both conditions require the total charging mass of coolant $M_{\text{fill}}$ in the LHP to persist within a certain range.
If the first condition is violated,
oscillations will occur during operation, as a result of $M_{\text{fill}}$ being too low  \cite{bai2009mathematical}.
Violation of the second condition implies that the coolant completely fills the LHP, which eliminates 
the condensation section in the condenser and results in a loss of capillary forces to drive the coolant circulation.

The heat-leakage limit for a certain operating temperature $T_{\text{op}}$ is identified by checking whether the coolant in the compensation chamber is saturated after reaching the operating temperature, namely $T_{\text{cc}} = T_{sat}(P_{\text{cc}})$ \cite{chen2021heat,qu2018numerical}.
In that case, any further increase of the input power $\dot{Q}_{\text{eff}}$ will lead to superheating of the coolant, and a consequent stop of coolant circulation in the loop. 

The activation of the heat-dissipation limit is assessed based on the criteria that $x_{\text{cond,outlet}} \geq 0$ and $T_{\text{cc}} = T_{sat}(P_{\text{cc}})$.
By definition, this limit is thus reached when the input heat power is no longer below the maximum heat dissipation rate of the LHP system, so that  both the condenser and the evaporator (and its integrated compensation chamber) are fully utilized.

The coolant limit is activated
when the temperature inside the loop becomes lower than the coolant's triple point temperature, or higher than the  coolant's critical temperature. 
That is when $T_{\text{triple}} \geq T_{\text{op}}$ or $T_{\text{op}} \geq T_{\text{crit}}$ \cite{zohuri2011heat}.

Finally, the application temperature limit requires that $T_{\text{op}} \leq T_{\text{app}}$, where the application temperature is set to {$T_{\text{app}}$ = 130 \celsius}.

\section{Discretization and numerical solution procedure }
\label{sec: solution} 
The preceding model equations are discretized by applying a finite-difference method on a 1D \textit{Cartesian} mesh with local grid size $\Delta s_i$ around each grid point $s_i$ in the flow direction (see Fig. \ref{fig01}). 
A second-order central scheme is used for the discretization of the diffusive term in the evaporator energy equation \ref{eq: thermal equation coolant in wick} and its boundary conditions, 
while a first-order \textit{Euler} backward scheme is used for the discretization of the convection terms (equations (\ref{eq: thermal equation coolant in wick}), (\ref{eq: energy balance single-phase regions condenser}), pressure gradients (equations (\ref{eq: momentum equation evaporator}) and (\ref{eq: momentum equation condenser})). 
All fluid properties are evaluated based on the local temperature and pressure at each grid point, using the database {\rmfamily\scshape Refprop} \cite{lemmon2007nist}.

\begin{figure}[htp] 
\centering
\includegraphics[width=0.8\textwidth, angle = 0]{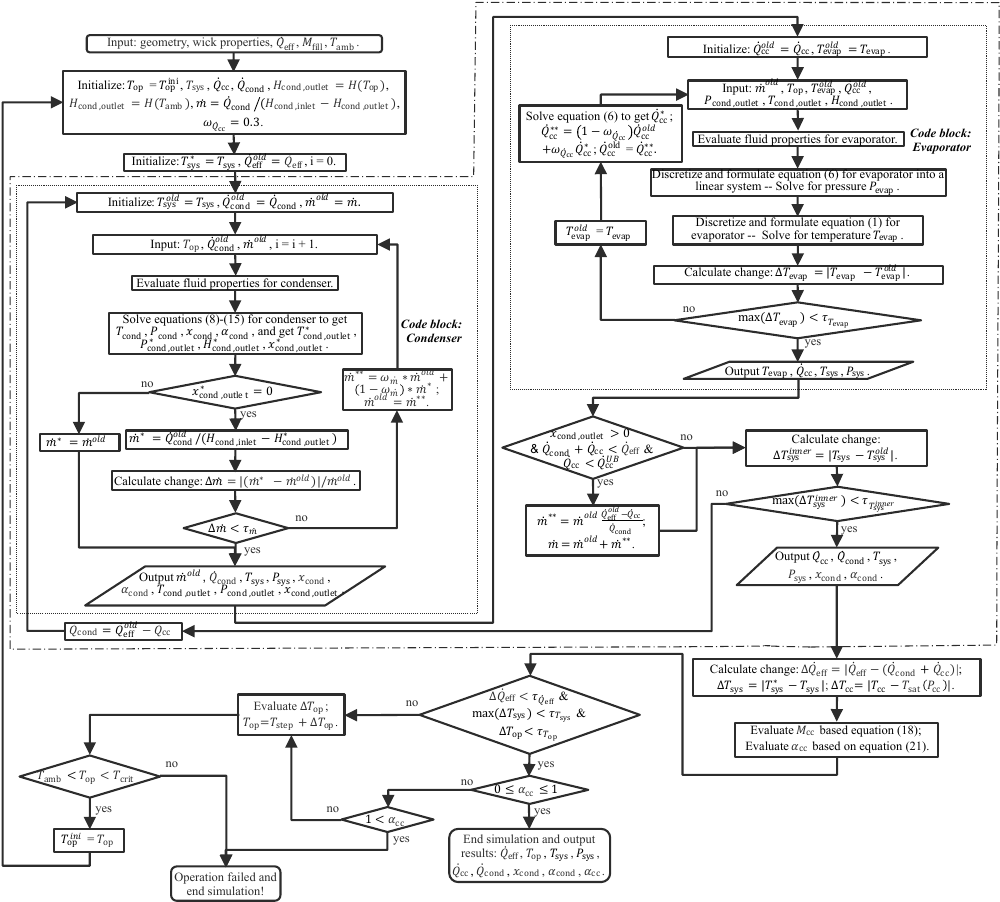}
\caption{Flowchart of solving the LHP system model with given $\dot{Q}_{\text{eff}}$ and $M_{\text{fill}}$.}
\label{fig02}
\end{figure}

Once the boundary conditions for the LHP (i.e., the ambient temperature $T_{\text{amb}}$, and the effective input heat power $\dot{Q}_{\text{eff}}$) are specified, the system of model equations is solved. 
As shown in Fig. \ref{fig02}, the set of equations is split into two blocks, which are solved sequentially through a fixed-point iteration scheme.
The first block of equations corresponds to the condenser model from section \ref{subsec: condenser model}. 
The second block consists of the model equations for the evaporator from section \ref{subsec: evaporator model}.

In the first block, the inlet conditions for the condenser ($P_{\text{cond,inlet}}$, $T_{\text{cond,inlet}}$) as well as the mass flow rate $\dot{m}$ are initialized from an initial guess or a previous iteration.
The fluid properties at each point $s_i$ in the condenser are evaluated using the last available temperature and pressure values.
First, the discretized energy equations (\ref{eq: energy balance single-phase regions condenser}) for the condenser are solved as a coupled linear system to obtain the temperature distribution in the condenser. 
Next, the discretized momentum equations (\ref{eq: momentum equation condenser}) are solved separately to get the pressure distribution in the condenser. 
As long as the vapor quality at the condenser outlet is zero, i.e.  $x_{\text{cond,outlet}} = 0$, a mass flow rate correction is performed based on the overall energy balance for the condenser 
$\dot{m} = \dot{Q}_{\text{cond}} / (H_{\text{cond,inlet}} - H_{\text{cond,outlet}})$, by using the new values for the pressures, temperatures and the related cooling rate in the condenser. 
For updating the mass flow rate through the condenser, under-relaxation with a factor {$\omega_{\dot{m}}$ = 0.5} is applied, until the relative change in the mass flow rate $\Delta \dot{m} = |\dot{m}^{*} - \dot{m}^{old}| / \dot{m}^{old}$ is less than {$\tau_{\dot{m}}={\rm 10^{-4}}$}. In case $x_{\text{cond,outlet}} >$ 0, the second block is first solved before adapting the mass flow further.

In the second block which evaluates the evaporator, the mass flow rate $\dot{m}$ and the inlet conditions for the evaporator ($P_{\text{cond,outlet}}$, $T_{\text{cond,outlet}}$) from the first block are used as inputs.
The unknown heat loss $\dot{Q}_{\text{cc}}$ from the evaporator 

\noindent is initially guessed.
The discretized momentum equations (\ref{eq: momentum equation evaporator}) are solved first to obtain the pressure in the evaporator.
Subsequently, the discretized energy equations (\ref{eq: thermal equation coolant in wick}) are solved to obtain the temperature. 
Due to the dependence of the fluid properties on the state variables, the nonlinearity of both equations is treated by evaluating all fluid properties based on the temperature and pressure from a previous iteration. 
It must be mentioned here that when the coolant liquid in the wick is superheated, the fluid properties are evaluated by assuming a saturation state. 
After the energy equation is solved, $\dot{Q}_{\text{cc}}$ is updated based on the new temperature $T_{\text{cc}}$ in a correction step with an under-relaxation factor of {$\omega_{\dot{Q}_{\text{cc}}}$ = 0.3}.
Convergence of the evaporator block for the specified mass flow rate and inlet conditions is reached when the maximum change of the temperature at each grid point, $\Delta T_{\text{evap}}$, is smaller than {$\tau_{T_{\text{evap}}}={\rm 10^{-8}}$ K}. 

It can be seen in Fig. \ref{fig02} that the  correction step for the mass flow rate, in case {$x_{\text{cond,outlet}}> $ 0}, is performed after the second code block only. This is done to prevent the mass flow rate to diverge. Indeed, the correction step in the condenser block cannot be used in this case as it would lead to an ever higher mass flow rate
as long as $\dot{m}(H_{\text{cond,inlet}} - H_{\text{cond,outlet}}) <  \dot{Q}_{\text{cond}}$, which in turn would lead to an higher vapor quality $x_{\text{cond,outlet}}$ and enthalpy $H_{\text{cond,outlet}}$ at the condenser outlet, and thus a lower enthalpy difference $(H_{\text{cond,inlet}} - H_{\text{cond,outlet}})$ over the condenser. Therefore, in case of {$x_{\text{cond,outlet}} >$ 0}, the mass flow rate is corrected  based on the overall energy balance for the LHP equation (\ref{eq: energy balance LHP}). 
It updates the mass flow rate via the factor $(\dot{Q}_{\text{eff}} - \dot{Q}_{\text{cc}})/\dot{Q}_{\text{cond}}$, which is no longer susceptible to the unstable mass flow correction for this case in the condenser block. Consequently, it leads to a more robust but slower convergence of the solution algorithm.
From a physics point of view, the second correction step takes into account that when {$x_{\text{cond,outlet}} >$ 0}, 
the condenser is not able to cool down the coolant into liquid because its maximum cooling capacity is reached. 
As such, the cooling capacity of the compensation chamber ($\dot{Q}_{\text{cc}}$) has to be utilized further. It should be noted that in our case $\dot{Q}_{\text{cc}}$ is relatively small compared to the cooling rate of the condenser. 

At the end of the core block, the inner iterations to update $\dot{Q}_{\text{cond}}$ based on the overall energy balance equation (\ref{eq: energy balance LHP}) continue 
until the maximum change in temperature for all points in the LHP system, $\Delta T_{\text{sys}}^{inner}$, is smaller than {$\tau_{T_{\text{sys}}^{inner}} = {\rm 1 \times 10^{-3}}$ K}. 
Upon termination of these iterations in the core block, all state variables will have converged to a value which matches the last available value for $T_{\text{op}}$, although $T_{\text{op}}$ might still be incompatible with the specified $M_{\text{fill}}$.
Therefore, $T_{\text{op}}$ itself is updated with a fixed increment $\Delta T_{\text{op}}$\footnote{To speed up  the convergence, we made use of some heuristic rules to choose $\Delta T_{\text{op}}$ in a manner that $0 \leq \alpha_{\text{cc}} \leq 1.0$ is achieved. These rules are omitted here for conciseness.}  
until convergence is reached for the entire system of equations, by checking whether the changes in $\dot{Q}_{\text{eff}}$, $T_{\text{op}}$,  
and the other temperatures are respectively below 
the prescribed tolerances
$\tau_{\dot{Q}_{\text{eff}}} = 1 \times 10^{-4}$ W,
$\tau_{T_{\text{op}}} = 1 \times 10^{-2}$ K
and
$\tau_{T_{\text{sys}}} = 1 \times 10^{-4}$ K.
We remark that the increment $\Delta T_{\text{op}}$ is only applied when
$\alpha_{\text{cc}} < 0$, as this indicates that the LHP should contain more coolant, hence a higher $T_{\text{op}}$.
That way we achieve that $0 \leq \alpha_{\text{cc}} \leq 1.0$ (within a tolerance of {$\tau_{\alpha_{\text{cc}}}={\rm 1\times 10^{-3}}$}).

\section{Experimental setup and test results}
\label{sec: experiment setup and results}
\subsection{Details of the manufactured LHP}
\label{subsec: manufactured lhp}
The designed LHP has an evaporator whose bottom surface is in contact with the heat source. The evaporator consists of a copper plate with dimensions of {30.0 $\times$ 34.5 mm} and a thickness of {0.5 mm}.
On top of this copper plate, four evenly-spaced beam-shaped fins are placed, having a width of {1.5 mm}, a length of {10 mm}, and a height of {5.5 mm}.
The grooves between the fins also have a width of {1.5 mm}.
The margins/offsets between the fins and the casing of the LHP along the main flow direction are {10.0 mm} and {14.0 mm}, before and behind the fins, respectively.
Perpendicular to the main flow direction, the margins/offsets are equal to {10.0 mm}.
Also, a space behind the fins (i.e. the accumulator) is present, which collects the vapor leaving the grooves.

The LHP's wick consists of a piece of copper foam with a cross-section of {10.5 $\times$ 10.0 mm} and a height of {4.5 mm}.
During the system assembly, this wick was pressed against the top surface of the fins to reduce the thermal contact resistance.
The wick characteristics are given in Table \ref{tab01}. 
The maximum pore size $r_{\text{pore}}$ of the wick is determined via the bubble point method \cite{dickenson1997filters}.
To measure 

\begin{table}[H]
\centering
\fontsize{7}{10}\selectfont
\caption{Measured wick parameters.}\label{tab01}
\begin{tabular}{ccccc}	
\hline
Parameter                & $r_{\text{pore}}$ (${\rm \mu m}$)      &  $K$ (${\rm m^2}$)       
                         & $\phi$ (-)                             &  $k_{\text{eff},l}$ ({$\rm W/m\cdot K$}) \\
\hline
Value                    & {${\rm 44.5\pm 0.13}$}                 & {${\rm (4.7\pm 0.2)\times 10^{-11}}$} 
                         & {${\rm (73.48\pm 0.05)\%}$}            & {${\rm 14.0\pm 0.2}$} \\
\hline
\end{tabular}
\end{table}
\noindent the porosity $\phi$, we used a vacuum saturation test\footnote{The vacuum saturation test is also called the \textit{Archimedes} method in literature \cite{wu2023effect,zhao2023r1234ze}.} \cite{feng2015hygric}. We determined the wick's permeability $K$ using the falling head method \cite{Fallingheadmethod}. 
Finally, the effective thermal conductivity of the wick (saturated with liquid water) is measured at Zhejiang University using a {\textit{TPS 2500 S}} device from \textit{Hot Disk} \cite{Hotdiskdevice}, which relies on the transient plane source method.

The casing around the LHP's evaporator consists of five polycarbonate (PC) plates of a thickness of {8.0 mm}.
This casing forms the compensation chamber above the wick, whose volume is {8.9 mL}.
We remark that the clearance between the casing and the wick was eliminated by inserting silicone rubber with a thickness of {1.0 mm} (before compression).
That way, leakage of the coolant from the compensation chamber into the grooves is avoided.
The clearance between the casing and the vapor collecting region behind the fins is filled with silicon rubber with a thickness of {5.0 mm}.

The condenser pipe of the LHP was a transparent Perfluoroalkoxy Alkane (PFA) tube  enabling us to measure the length of the two-phase flow region optically.
Finally, we remark that degassed water is the coolant used during the experimental campaigns.

\subsection{Details of the experimental setup}
\label{subsec: experimental setup}
A schematic diagram of our experimental setup for the LHP is shown in Fig. \ref{fig01} (not to scale). 
The setup is positioned horizontally on an optical table, in order to mimic the LHP operation in a regular laptop.

The first item in our setup is the heating simulator, which serves as an emulator of a real chip with an active heated area {$A_{\text{heat}}$ = 1 ${\rm cm^2}$}.
It consists of a single block of oxygen-free copper, which distributes the heat generated by four thick-film resistors \cite{Tabletheatingelement} uniformly towards the LHP’s evaporator.
The four thick-film resistors are symmetrically mounted on the sides of the copper heating block using a thermal paste with a thermal conductivity of {5.0 ${\rm W/m\cdot K}$}.
They are connected in series to a DC power supply (\textit{Agilent N5751A}), which allows controlling the heat load towards the evaporator. 
Five sides of this heating system are surrounded by insulating material (Teflon and Peek) with thermal conductivities lower than {0.3 W/m$\cdot$K} to limit the heat loss to the ambient. 
As indicated in Fig. \ref{fig01}, two pieces of steel plate are used as part of the fixture to clamp the Teflon and PEEK material together.
Teflon and PEEK were selected to withstand the high temperature of the copper heating block: their glass transition temperature lies above {110 \celsius}.
To further reduce the heat loss, an insulating air gap of more than {2 mm} is ensured between the fixture, the outer surface of the resistors, and the copper heating block.
In addition, the heater is insulated by two layers of ceramic fiber plate (thickness {10.0 mm} with thermal conductivity of {0.085 ${\rm W/m\cdot K}$}) and two layers of the polymer foil (thickness {3.0 mm}, each with a thermal resistance of {0.083 ${\rm m^2\cdot K/W}$}) during all tests, as indicated in Fig. \ref{fig01}.

One absolute pressure transducer (type \textit{PMP 4070D}) is used to measure the total pressure at the top of the compensation chamber.
Eight calibrated \textit{T}-type thermocouples are used to measure the temperature at different positions in the LHP and setup, as indicated in Fig. \ref{fig01}. 
The thermocouples provide: 
\begin{itemize}
    \item the temperature $T_{\text{cu,bot}}$ at the bottom of the copper heating block and the temperature $T_{\text{st,bot}}$ at the bottom of the steel plate,
    \item the coolant temperatures $T_{\text{cc,top}}$, $T_{\text{cc,mid}}$ and $T_{\text{cc,bot}}$ at the top, middle and bottom of the compensation chamber, respectively. We note that $T_{\text{cc,bot}}$ corresponds to the coolant temperature at the wick's cold side facing the compensation chamber,
    \item the coolant temperature $T_{\text{groove}}$ in the middle of the fin grooves, which we consider the operating (or vaporization) temperature $T_{\text{op}}$. The thermocouple measuring $T_{\text{groove}}$ is located at a distance of {1 cm} from the condenser inlet,
    \item the coolant temperature $T_{\text{cond,inlet}}$ in the vapor collection region near the inlet of the condenser,
    \item the ambient temperature $T_{\text{amb}}$.
\end{itemize}
 
The signals from the pressure transducer and thermocouples are acquired using \textit{LabVIEW}  and a combination of a National Instruments data acquisition system and an \textit{Agilent 34970A} data logger with a sampling frequency of {1 Hertz}.
The length of the two-phase flow region in the condenser, $L_{\text{tp}}$, is measured by direct visualization.

As indicated in Fig. \ref{fig01}, the setup has a degassing valve, a charging valve, and a control valve. 
At the start of the experiment, the degassing valve is briefly opened to extract the air from the LHP loop via a vacuum pump, which maintains a vacuum pressure of {$P_{\text{vcu}}$ = -97.7 kPa}.
Then, the charging valve between the coolant reservoir and the LHP is opened to let the coolant flow into the LHP loop.
During the charging of the LHP, the control valve remains closed.
That way, the coolant is forced to flow through the condenser before reaching the evaporator, preventing any non-condensable gasses (NCG) in the loop from accumulating just in the CC.
After closing the charging valve, the control valve is opened, allowing the coolant to redistribute itself passively. 

\subsection{Experimental test results}
\label{subsec: experimental result}
\begin{figure}[htp] 
\centering
\includegraphics[width=0.8\textwidth, angle = 0]{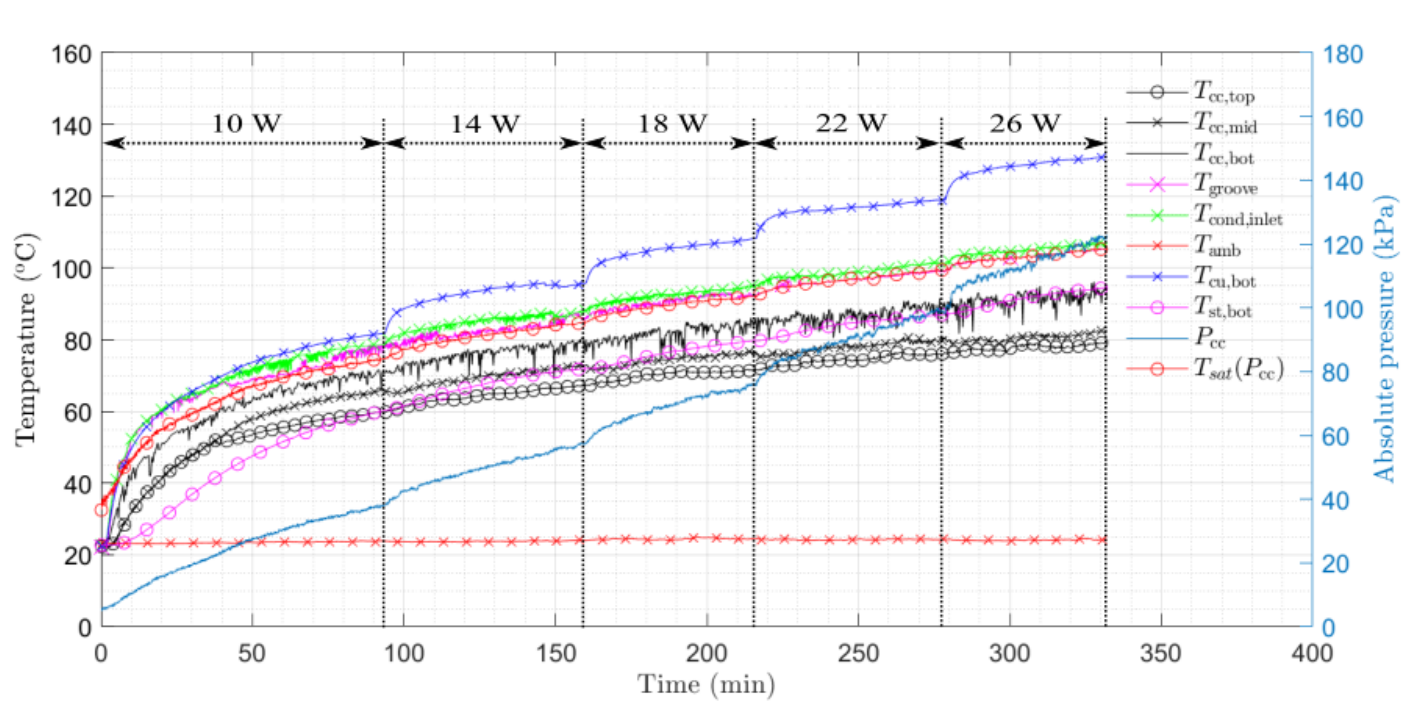}
\caption{Experimental test result of the LHP with a charging mass of $M_{\text{fill}}$ = {0.00997 kg} and stepwise increasing total input power {$\dot{Q}_{\text{tot}}$ = {10-26} W}.}
\label{fig03}
\end{figure}

To validate the numerical model in section \ref{sec: lhp model}, the LHP's temperature response is measured for five different step changes of the total input power $\dot{Q}_{\text{tot}}$, as shown in Fig. \ref{fig03}.
At the beginning of the test, the total input power $\dot{Q}_{\text{tot}}$ was increased from {0} to {10.0 W}.
At each consecutive step, after a quasi-steady reg-

\noindent ime was reached, $\dot{Q}_{\text{tot}}$ was further increased by {4.0 W}.
The test is performed using a constant charging mass of {$M_{\text{fill}}$ = 0.00997 kg}.

Fig. \ref{fig03} reveals that, once the input power is turned on ({$t >$ 0}), the temperature in the grooves ($T_{\text{groove}}$) and condenser inlet ($T_{\text{cond,inlet}}$) begins to increase over time with almost no delay. 
The reason is that the copper material in the heating block and the evaporator has a high thermal diffusivity (around {${\rm 1.11 \times 10^{-4}}$ ${\rm m^2/s}$}) so that the change in temperature of the heater is quickly transferred to the LHP's bottom plate and fins, and hence the coolant in the grooves.

After three minutes, both $T_{\text{groove}}$ and $T_{\text{cond,inlet}}$ (the condenser inlet temperature) exceed the saturation temperature $T_{\text{sat}}(P_{\text{cc}})$.
The temperature in the compensation chamber ($T_{\text{cc,bot}}$) rises as well over time, although with some delay. 
This delay can be attributed to the wick's low thermal diffusivity, which can be estimated as {$k_{\text{eff,l}}/[\phi \rho_{l} c_{p,l} + (1-\phi) \rho_{cu} c_{cu}]$ = ${\rm 3.50 \times 10^{-6}}$ ${\rm m^2/s}$}. 
As such, the heat transfer from the coolant in the grooves to the wick's cold side happens more slowly so that the increase of $T_{\text{cc,bot}}$ lags behind that of $T_{\text{groove}}$. 

The LHP is considered to operate in a (quasi-)steady regime when the temperature $T_{\text{st,bot}}$ fluctuates with less than {{5 \celsius} per hour}. 
The other temperature signals are less suitable for identifying the quasi-steady regime. 
As we explained, the temperature response observed for $T_{\text{cu,bot}}$ is, for instance, very fast. 
On the other hand, the response at the bottom side of the insulation material is extremely slow due to the low thermal diffusivity of the heater's fixture and its insulation material, which account for about {66\%} of LHP's total thermal capacity. 

\subsection{Experimental estimation of the effective input power, mass flow rate, and condenser heat dissipation rate}
\label{subsec: effective input power}
As the experimental results are to be compared with our LHP model for the same effective input power $\dot{Q}_{\text{eff}}$ from equation (\ref{eq: energy balance at wick/groove boundary}),
we estimated the latter for each test step in Table \ref{tab02}. 
 $\dot{Q}_{\text{tot}}$ is calculated using the following energy balance: 
\begin{equation}
\dot{Q}_{\text{eff}} = \dot{Q}_{\text{tot}} - \dot{Q}_{\text{loss}} - C_{\text{LHP}}\frac{dT_{\text{LHP}}}{dt} \,,
\end{equation}
in which the heat loss from the heater towards the ambient $\dot{Q}_{\text{loss}}$ is 
\begin{equation}
\dot{Q}_{\text{loss}} =
\frac{T_{\text{cu,bot}} - T_{\text{amb}}}{R_{\text{heater}}} \,,
\end{equation}
with $R_{\text{heater}}$ being the thermal resistance between the heating element and the ambient.
The determination of the thermal resistance $R_{\text{heater}}$ is reported in \ref{app: thermal resistance from copper heating block to ambient}. 
Namely, we assume that the heat loss occurs via perpendicular conduction through the bottom and sides of the heating simulator fixture and the surrounding insulation material. 
Passive convection at the insulation material's vertical surface is modelled by the same passive heat transfer coefficients as used in \ref{app: thermal resistance from coolant in cc to ambient}.
In the evaluation, we assume that the entire heater has a uniform temperature equal to the measured temperature at its bottom $T_{\text{cu,bot}}$.

The change in internal energy of the entire LHP during transient operation is estimated by adding the thermal capacities of all the LHP components and materials into a single lumped capacity 
$C_{\text{LHP}}$, which we find to be close to {462 J/K}.
The rate of change of the average temperature of the entire LHP, on the other hand, is approximated by averaging the time derivatives of the measured temperatures at the bottom of the heater, 
the grooves and the top of the compensation chamber: 
\begin{equation}
\left( \frac{dT_{\text{LHP}}}{dt} \right)  \approx
\frac{1}{3}
\left(
\frac{dT_{\text{cu,bot}}}{dt} + \frac{dT_{\text{groove}}}{dt} + \frac{dT_{\text{cc,top}}}{dt} 
\right) 
\,.
\end{equation} 
The time derivatives on the right-hand side of the former equation are approximated by the central finite difference of each measured temperature signal over a discrete time interval of {180 s}.

To determine the mass flow rate in the experiments, the velocity $u_{l}$ of the liquid phase in the condenser is measured via the bubble velocity method since $\dot{m} = A_{\text{in}} u_{l} \rho _{l}$.
The bubble velocity method relies on a measurement of the time interval $\Delta t$ in which the liquid plugs in the condenser travel over a certain distance $\Delta s$ to obtain $u_{l} \simeq \Delta s/ \Delta t$.
Of course, the method requires that the liquid droplets formed at the condenser wall agglomerate into larger plugs, which occupy the whole cross-sectional area $A_{\text{in}}$ of the condenser pipe.
In our experiments, this is effectively the case because the Bond number $Bo =  D_{\text{in}} / \sqrt{{\sigma}/{[g(\rho_l - \rho_v)]}}$ is much smaller than the critical bond number {$Bo_{\text{crit}}$ $\simeq$ 2} \cite{zhang2008advances,spinato2014thermo}.
Indeed, for water at {100 \celsius}, the critical diameter {$D_{\text{crit}}$ = 5 mm} corresponding to {$Bo_{\text{crit}}$ = 2} is almost twice as large as the actual inner diameter of the condenser pipe in our setup ({$D_{\text{in}}$ = 2.76 mm}).
Moreover, in our setup, the optical tracing of the liquid slugs is facilitated due to the presence of plugs of NCG in the condenser pipe.
These plugs of NCG, which appear between the

\begin{table}[H]
\centering
\fontsize{7}{10}\selectfont
\setlength\tabcolsep{2.0pt}
\caption{Evaluation of the effective input power and heat loss in the condenser.}\label{tab02}
\begin{tabular}{p{0.6cm}p{0.5cm}p{1.3cm}p{0.9cm}p{0.9cm}p{0.7cm}p{0.7cm}p{1.4cm}p{1.3cm}p{1.3cm}p{0.7cm}}
\hline
\multirow{2}{0.5cm}{Data point}  &   $\dot{Q}_{\text{tot}}$               & $T_{\text{op}}$                 & $T_{\text{cu,bot}}$             &   $\dot{Q}_{\text{sys,ine}}$           & $\dot{Q}_{\text{loss}}$    & $\dot{Q}_{\text{eff}}$          &   $u_{l}$                               & $\dot{m}$                         & $\dot{Q}_{\text{cond}}$         & $V_{\text{NCG}}$  \\
                                 &   (W)                                  & (\celsius)                      &  (\celsius)                      &  (W)                                  &  
(W)                        &  (W)                            &   ({${\rm 10^{-4}}$ m/s})               & ({${\rm 10^{-3}}$ g/s})           &  (W) 
                          & (mL)
\\
\hline
01                              &               10                        & 77.6 $\pm$ 2.8                  & 80.8                            &               0.92                      & 1.58                            & 7.50                            &               3.5 $\pm$ 0.6             & 2.1 $\pm$ 0.8                     & 5.2 $\pm$ 2.0 
                          & 3.24 \\

02                              &               14                        & 84.3 $\pm$ 2.9                  & 95.5                            &               0.19                      & 2.00                            & 11.81                           &               6.4 $\pm$ 1.0             & 3.8 $\pm$ 1.5                     & 9.6 $\pm$ 3.7 
                          & 2.63 \\

03                              &                18                       & 93.2 $\pm$ 3.1                  & 107.0                           &                0.34                     & 2.34                            & 15.32                           &                8.1 $\pm$ 0.5            & 4.8 $\pm$ 1.7                     & 12.3 $\pm$ 4.4 
                          & 2.32 \\  

04                              &               22                        & 98.7 $\pm$3.2                   & 118.1                           &               0.40                      & 2.67                            & 18.93                           &               11.1 $\pm$ 0.9            & 6.6 $\pm$ 2.4                     & 16.9 $\pm$ 6.1
                          & 2.01 \\  

05                              &               26                        & 105.2 $\pm$3.4                  & 130.4                           &               0.66                      & 3.03                            & 22.31                           &               13.2 $\pm$ 2.0            & 7.8 $\pm$ 3.0                     & 20.1 $\pm$ 7.6 
                          & 1.81 \\  
\hline
\end{tabular}
\end{table}
\noindent liquid plugs, not only move at the same average velocity $u_{l}$ but are also clearly visible and easy to capture on camera.

To evaluate the heat dissipation rate in the condenser $\dot{Q}_{\text{cond}}$ in equation (\ref{eq: energy balance condenser}), the enthalpy at the condenser inlet $H_{\text{cond,inlet}} = H_{\text{wick/groove}}$ is evaluated at temperature $T_{\text{groove}}$, while the enthalpy at the condenser outlet $H_{\text{cond,outlet}} = c_{P,l} T_{\text{cond,outlet}}$ is evaluated by assuming $T_{\text{cond,outlet}} = T_{\text{amb}}$. The latter assumption is reasonable considering that the coolant mass flow rate $\dot{m}$ is rather low, and the subcooled section is long enough to cool the liquid down to the ambient temperature at the pipe outlet.

With the previous methods, $\dot{Q}_{\text{eff}}$, $\dot{m}$ and $\dot{Q}_{\text{cond}}$ are determined for each experimental data point in Table \ref{tab02}. The associated uncertainty is evaluated with a safety factor of {3.0}.
Table \ref{tab02} also includes the volume of non-condensable gasses $V_{\text{NCG}} $ during each test step. These values are calculated from the measured value for $M_{\text{fill}}$ and the calculated value for $M_{\text{wick}}$, $M_{\text{cond}}$, $M_{\text{groove}}$, via equation (\ref{eq: mass in cc}), as $\alpha_{cc}=0$ for all five data points.

\section{Results and discussion}
\label{sec: results and discussion}
\subsection{Model calibration and validation}
\label{subsec: model calibration and validation}
\subsubsection{Calibration of the heat transfer coefficients for the condenser}
\label{subsubsec: heat transfer coefficient calibration}
In order to calibrate the heat transfer coefficients $h_{\text{in}}$ and $h_{\text{out}}$ at the inside and outside of the condenser, 
we determine the two tuning parameters $\xi _{\text{in}}$ and $\xi _{\text{out}}$ from section \ref{subsec: condenser model}. A high value of $\xi _{\text{in}}=10.0$ is set because we observe in our experiments that the contact angle between the water and the PFA material of the condenser pipe exceeds {${\rm 90^\circ}$}.
Therefore, dropwise condensation instead of filmwise condensation is induced inside the condenser.
As a result, the actual heat transfer coefficient is expected to be an order of magnitude higher than predicted by the chosen correlation \cite{carey2020liquid}.
After the determination of $\xi _{\text{in}}$, the value of $\xi _{\text{out}}$ is determined through a least-square fitting procedure.
The underlying principle is to match the value of the two-phase flow section length $L_{\text{tp}}^{N}$ predicted by our numerical condenser model as closely as possible to its experimental value $L_{\text{tp}}$, 
minimizing the sum of their squared differences for the five data points given in Table \ref{tab03}. 
This fitting procedure results in the value $\xi_{\text{out}} = 1.41$, when the directly measured $T_{\text{op}}$ and the derived $\dot{Q}_{\text{cond}}$ are used as inputs for our condenser model. 

\begin{table}[H]
\centering
\fontsize{7}{10}\selectfont
\setlength\tabcolsep{2.0pt}
\caption{Validation of the condenser and evaporator model.}
\label{tab03}
\begin{tabular}{p{0.6cm}p{0.5cm}p{0.7cm}p{0.7cm}p{0.7cm}p{0.7cm}p{0.8cm}p{1.5cm}p{1.5cm}p{0.7cm}p{0.7cm}p{0.7cm}p{0.7cm}}
\hline
\multirow{5}{0.5cm}{Data point} & \multicolumn{6}{c}{Experimental data} & \multicolumn{2}{c}{Heat transfer coefficient }                     & \multicolumn{4}{c}{Condenser and evaporator}
\\
                                & \multicolumn{6}{c}{}                  & \multicolumn{2}{c}{calibration}     & \multicolumn{4}{c}{model validation}
\\
\cline{2-13}
                                & $\dot{Q}_{\text{tot}}$                & $T_{\text{op}}$                       & 
$\dot{Q}_{\text{eff}}$          & $\dot{Q}_{\text{cond}}$               & $L_{\text{tp}}$                       & 
$T_{\text{cc,bot}}$             
& $L_{\text{tp}}^{N}$                                   & $L_{\text{tp}}^{N}$  
& $\dot{Q}_{\text{cond}}^{N}$   & $L_{\text{tp}}^{N}$                   & $T_{\text{cc}}^{N}$                   & $M_{\text{cond}}^{N}$           
\\
                                &                                       &                                       & 
                                &                                       &                       
                                &              
& ($\xi_{\text{out}}$=1.00)                            & ($\xi_{\text{out}}$=1.41) 
                &                               &                       &                         &             
\\
                                &  (W)                                  & (\celsius)                            & 
(W)                             &  (W)                                  & (cm)                                  & 
(\celsius)                      
&  (cm)                                                          &  (cm)              
&  (W)                          & (cm)                                  & 
(\celsius)                      & (g)                                   
\\
\hline
01                              & 10                                    & 77.6 $\pm$ 2.8                        & 
7.50                            & 5.2 \,\,$\pm$ 2.0                         & 28.0 $\pm$ 2.0                        & 
70.7 \,\,$\pm$ 2.8                
& 30.76                                                                 & 24.15 
& 6.31                          & 30.52                                 & 
71.49                           & 4.13                                   
\\

02                              & 14                                    & 84.3 $\pm$ 2.9                        & 
11.81                           & 9.6 \,\,$\pm$ 3.7                         & 39.0 $\pm$ 1.0                        & 
78.6 \,\,$\pm$ 3.0                
& 49.11                                                                 & 38.50 
& 10.49                         & 43.67                                 & 
76.16                           & 3.39                                   
\\

03                              & 18                                    & 93.2 $\pm$ 3.1                        & 
15.92                           & 12.3 $\pm$ 4.4                        & 44.0 $\pm$ 2.0                        & 
83.8 \,\,$\pm$ 3.2                
& 52.67                                                                 & 41.87 
& 14.43                         & 50.11                                 & 
82.56                           & 3.03                                   
\\  

04                              & 22                                    & 98.7 $\pm$3.2                         & 
18.93                           & 16.9 $\pm$ 6.1                        & 52.0 $\pm$ 2.0                        & 
88.4 \,\,$\pm$ 3.3                
& 65.50                                                                 & 52.08 
& 17.33                         & 54.41                                 & 
86.25                           & 2.79                                   
\\  

05                              & 26                                    & 105.2 $\pm$3.4                        & 
22.31                           & 20.1 $\pm$ 7.6                        & 52.5 $\pm$ 2.5                        & 
93.2 \,\,$\pm$ 3.4                
& 70.82                                                                 & 55.61 
& 20.59                         & 57.87                                 & 
90.60                           & 2.60                                  
\\  
\hline
\end{tabular}
\end{table}

In the literature, it is rather uncommon to calibrate the available heat transfer coefficient correlations, as the condenser model is usually not validated separately. 
Nevertheless, the necessity of this calibration becomes evident from the fact that for $\xi _{\text{in}} = \xi _{\text{out}}=1$, the relative difference between $L_{\text{tp}}^{N}$ and $L_{\text{tp}}$ varies between $9.9\%$ and $34.9\%$.
On the contrary, after determining of $\xi _{\text{in}}$ and $\xi _{\text{out}}$, this relative difference shrinks to $0.8\%$ and $13.8 \%$, respectively.
We thus conclude that the former calibration of the condenser model is essential for reducing the discrepancies between the measured and predicted temperatures.

\subsubsection{LHP model validation}
\label{subsubsec: model validation and discrepancy}
As a first validation step, we compute $T_{\text{cc}}$ and $L_{\text{tp}}$ using the calibrated condenser model and evaporator model together, and we compare them to the experimental data.
To this end, we use the measured input power $\dot{Q}_{\text{eff}}$ and operating temperature $T_{\text{op}}$ as model inputs. 
The results of this validation step are presented in the column of $L^{N}_{\text{tp}}$ in Table \ref{tab03}. 
It can be deduced that the relative difference between the measured and numerically calculated values for $L_{\text{tp}}$ varies between $4.6\%$ and $13.9\%$.
Furthermore, the relative difference between the measured and numerically calculated values of $T_{\text{cc}}$ lies within the range of $1.0\%$ to $3.1\%$. 

To put this comparison into perspective, we note that the adopted \textit{Morgan} correlation has an uncertainty of {5\%} \cite{morgan1975overall}, but also has been reported to have uncertainty above {13\%} for the cylinder with small diameters or the \textit{Rayleigh} number $R_a$ is small \cite{davies2000gaseous}.
Given intrinsic modelling errors, the former validation confirms the present model's capability to reproduce the average temperatures and the main features of the two-phase heat transfer.

\begin{figure}[htp] 
\centering
\includegraphics[width=0.8\textwidth, angle = 0]{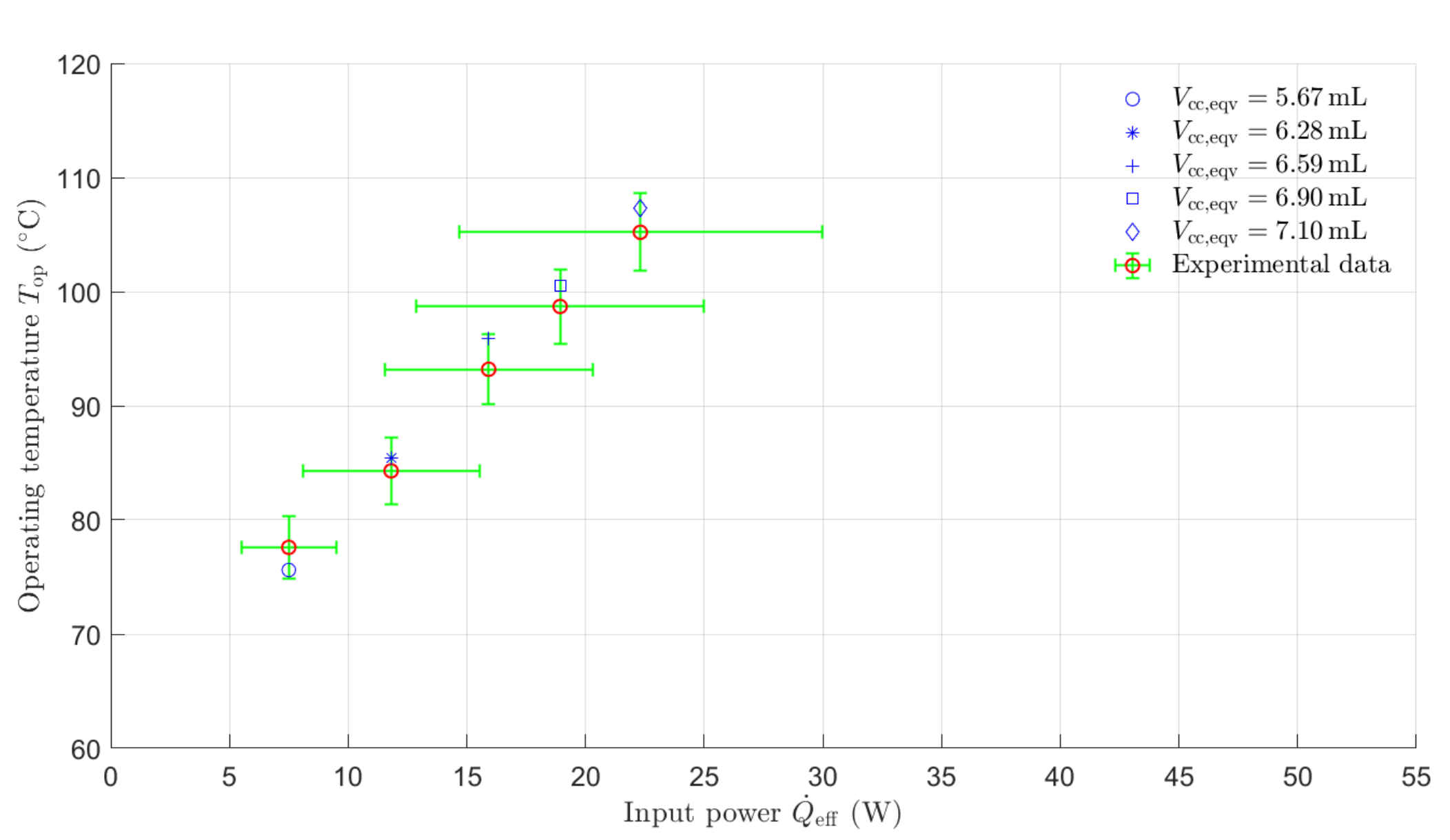}
\caption{The comparison between the predicted and measured LHP operating temperature with charging mass of {$M_{\text{fill}}$ = {0.00997 kg}} (the effect of NCG included).}
\label{fig04}
\end{figure}

For validating the entire LHP model, the operating temperature $T_{\text{op}}^{N}$ predicted by the LHP model is compared with its measured value $T_{\text{op}}$. 
This comparison is plotted in Fig. \ref{fig04}. All the relative differences fall within $1.9\%$ and $3.5\%$. 
It shows that the entire LHP model accurately predicts the LHP's operating point and performance.
We clarify that the results in Fig. \ref{fig04} are obtained

\noindent using the measured input power $\dot{Q}_{\text{eff}}$ and the measured charging mass $M_{\text{fill}}$ as model inputs,
in accordance with the flowchart from Fig. \ref{fig02}. 
Besides this, the actual geometrical parameters of the LHP from Table \ref{tab_app_01} (see \ref{app: thermal resistance from coolant in cc to ambient}) are used together with the estimated volume $V_{\text{NCG}}$\footnote{The method of estimating $V_{\text{NCG}}$ is similar to the method reported in \cite{anand2022novel}.} occupied by non-condensable gasses during the experimental tests (see Table \ref{tab02}), resulting in an equivalent compensation chamber volume $V_{\text{cc,eqv}}$. 
Ignoring the presence of non-condensable gasses (so that $V_{\text{NCG}}=0$) would lead, in our case, to an underestimated prediction of operating temperature with a mean discrepancy of {$24.97$ \celsius}. This corresponds to the NCG's effect on LHPs system performance reported in \cite{anand2022novel}.
Despite this observation, the presence of non-condensable gasses is mostly ignored by previous validation of numerical models \cite{chuang2002comparison,launay2008analytical,adoni2009effects,bai2009mathematical}.
Likely, the reason is that most of the experimental results are obtained from LHPs with little amount of NCGs. On the one hand, this can be attributed to the well-chosen vacuum pumps and well-designed charging procedure that can remove most of the NCGs in the loop. On the other hand, these LHPs are fabricated largely by welding of metallic components or by strong mechanical connection, so that these LHPs are sealed well and the entrance of non-condensable gasses in the loop is prevented \cite{singh2007miniature,wang2014experimental,mitomi2014long,anand2018experimental,anand2022novel,maydanik2022visual,zhao2023r1234ze,zhang2023performance}.

\subsection{The predicted operating characteristics of the LHP system and its subsystem}
\label{subsec: operating characteristics}
\subsubsection{System performance curves and the influence of the charging mass}
\label{subsubsec: performance curves}
\begin{figure}[htp] 
\centering
\includegraphics[width=1.0\textwidth, angle = 0]{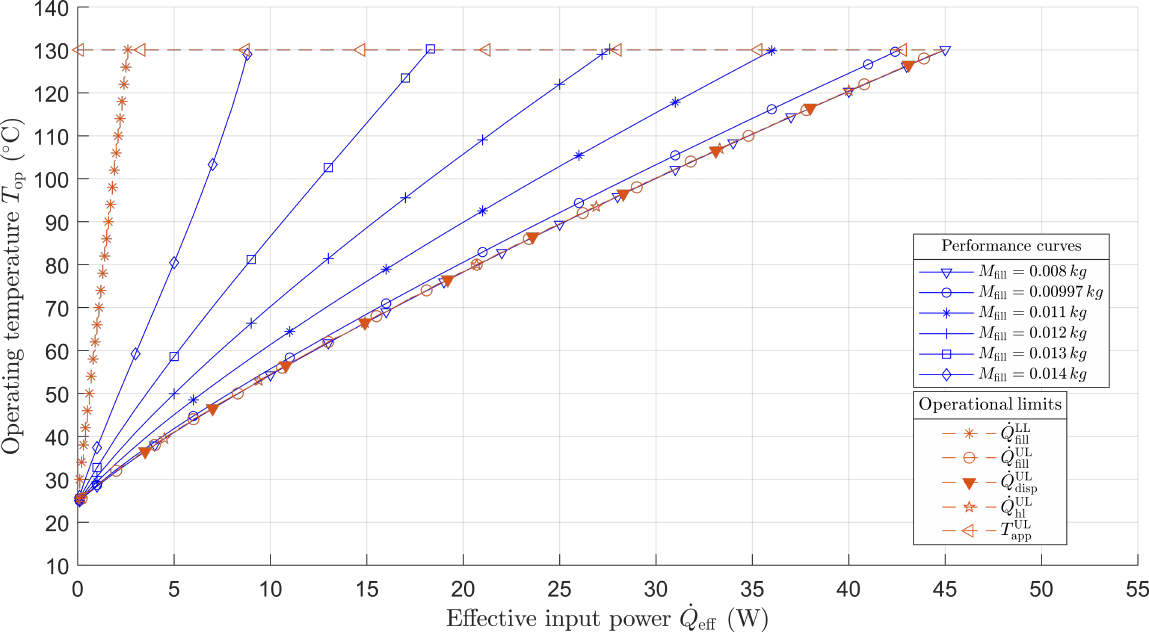}
\caption{The predicted LHP system performance curves with various charging mass of coolant and the operational envelope formed by the operational limits for current setup.}
\label{fig05}
\end{figure}

Using our LHP model, we have determined the performance curves of the LHP in our experimental setup for six different charging masses within the range (0.008, 0.014) kg. 
These six performance curves are shown in Fig. \ref{fig05}. They disp- 

\noindent lay three main trends.
First, a higher effective input power $\dot{Q}_{\text{eff}}$ always results in a higher operating temperature $T_{\text{op}}$. 
For example, if the charging mass equals {$M_{\text{fill}}=0.012$ kg}, 
$T_{\text{op}}$ increases from {70.2 \celsius} to {105.7 \celsius} upon an increase of $\dot{Q}_{\text{eff}}$ from {10 W} to {20 W} (see the curve indicated by the $+$ markers).
Secondly, all the curves feature a quasi-linear relation between $T_{\text{op}}$ and $\dot{Q}_{\text{eff}}$. 
Thirdly, the slope of each performance curve becomes smaller with smaller charging mass $M_{\text{fill}}$.
As all performance curves intersect at the origin where $\dot{Q}_{\text{eff}}=0$, this implies that a lower operating temperature $T_{\text{op}}$ is achieved by decreasing the charging mass $M_{\text{fill}}$ for any given input power $\dot{Q}_{\text{eff}}$.

\begin{figure}
     \centering
     \begin{subfigure}[b]{0.48\textwidth}
         \centering
         \includegraphics[width=\textwidth, angle = 0]{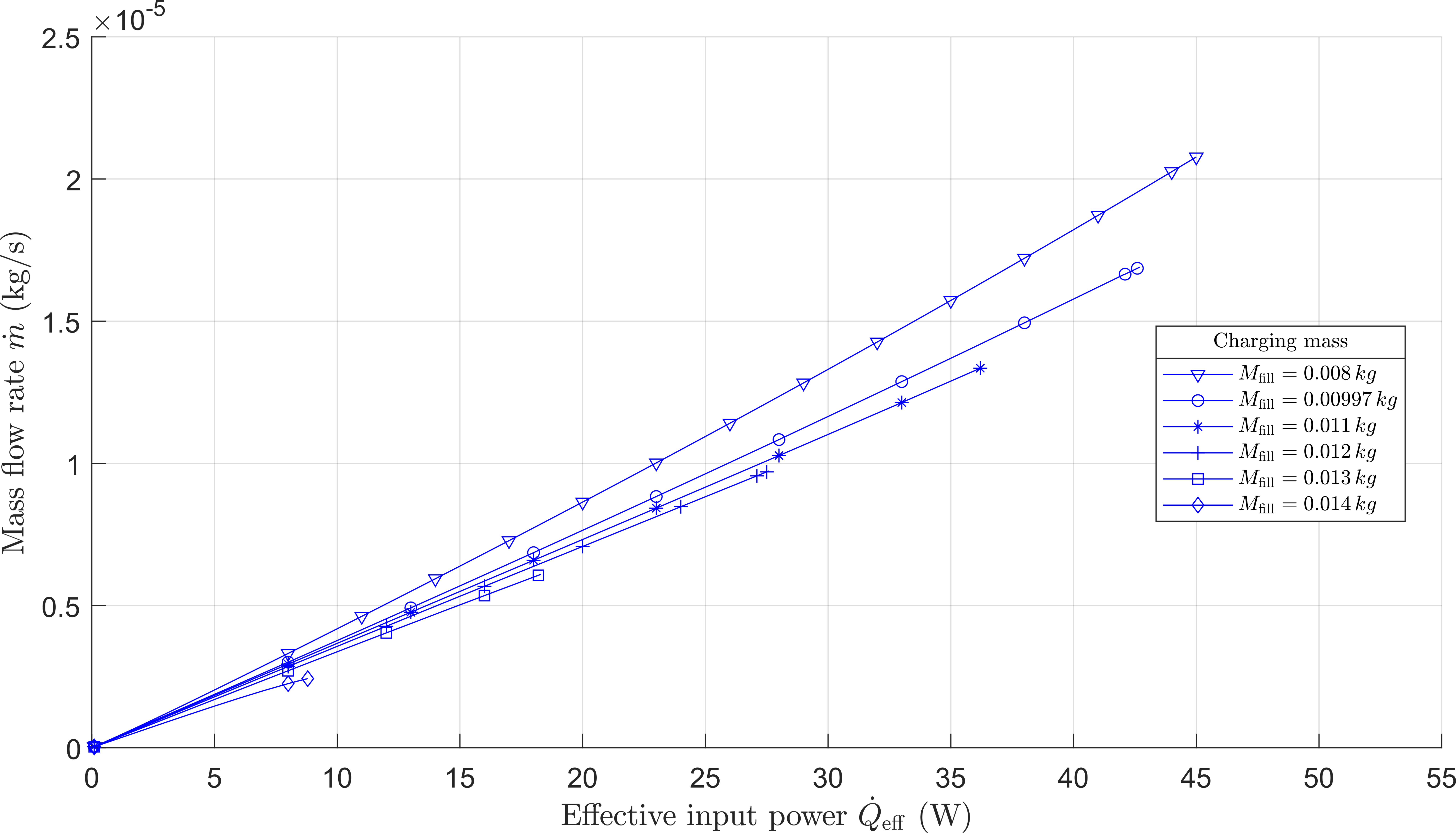}
         \caption{Coolant mass flow rate in the loop.}
         \label{fig06a}
     \end{subfigure}
      \hspace{0.5mm}
           \begin{subfigure}[b]{0.48\textwidth}
         \centering
         \includegraphics[width=\textwidth, angle = 0]{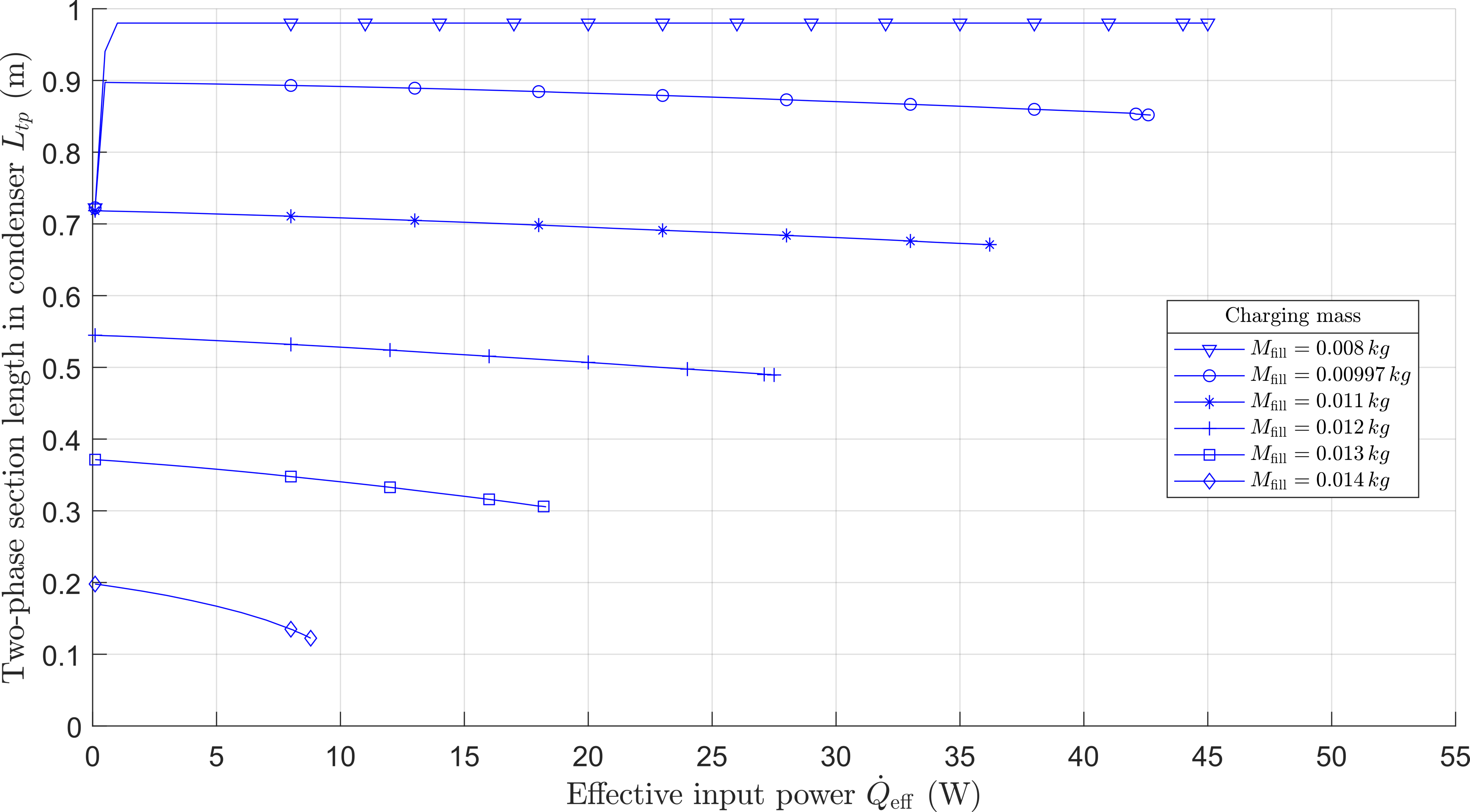}
         \caption{Two-phase flow section length in condenser.}
         \label{fig06b}
     \end{subfigure}
      \hspace{0.5mm}
           \begin{subfigure}[b]{0.48\textwidth}
         \centering
         \includegraphics[width=\textwidth, angle = 0]{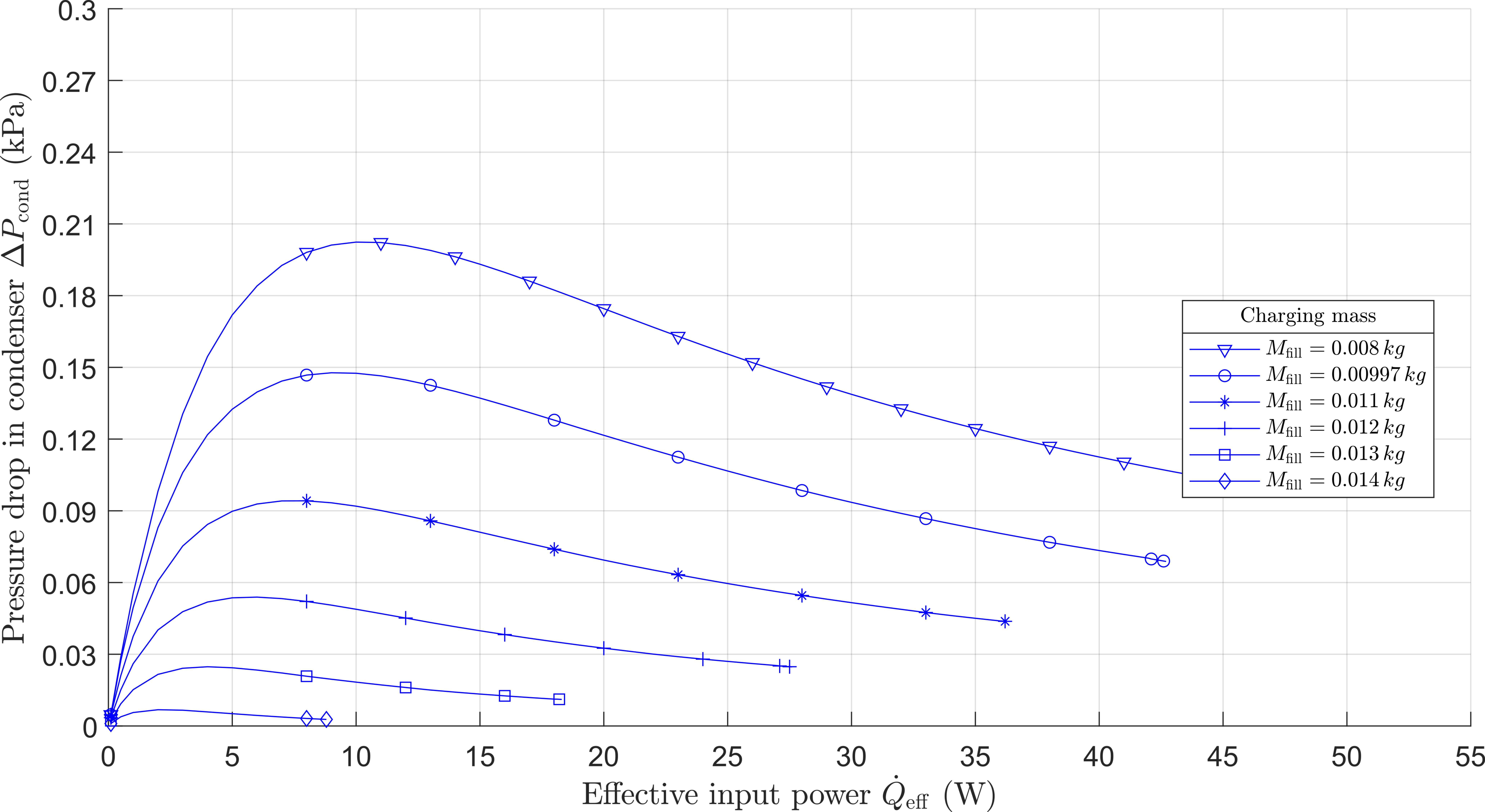}
         \caption{Coolant Pressure drop flowing through condenser.}
         \label{fig06c}
     \end{subfigure}
      \hspace{0.5mm}
     \begin{subfigure}[b]{0.48\textwidth}
         \centering
         \includegraphics[width=\textwidth, angle = 0]{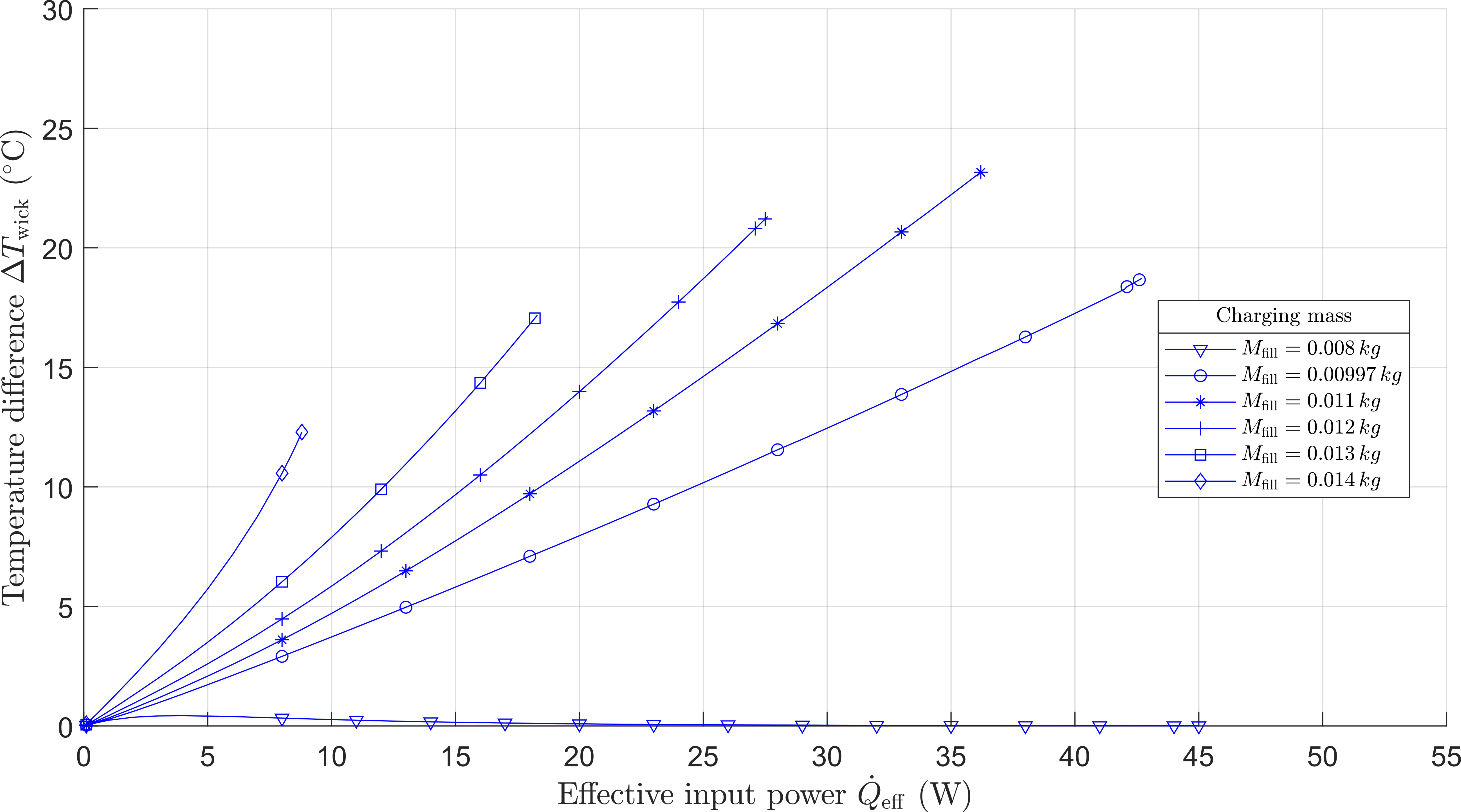}
         \caption{Temperature difference across the wick.}
         \label{fig06d}
     \end{subfigure}
      \hspace{0.5mm}     
     \begin{subfigure}[b]{0.48\textwidth}
         \centering
         \includegraphics[width=\textwidth, angle = 0]{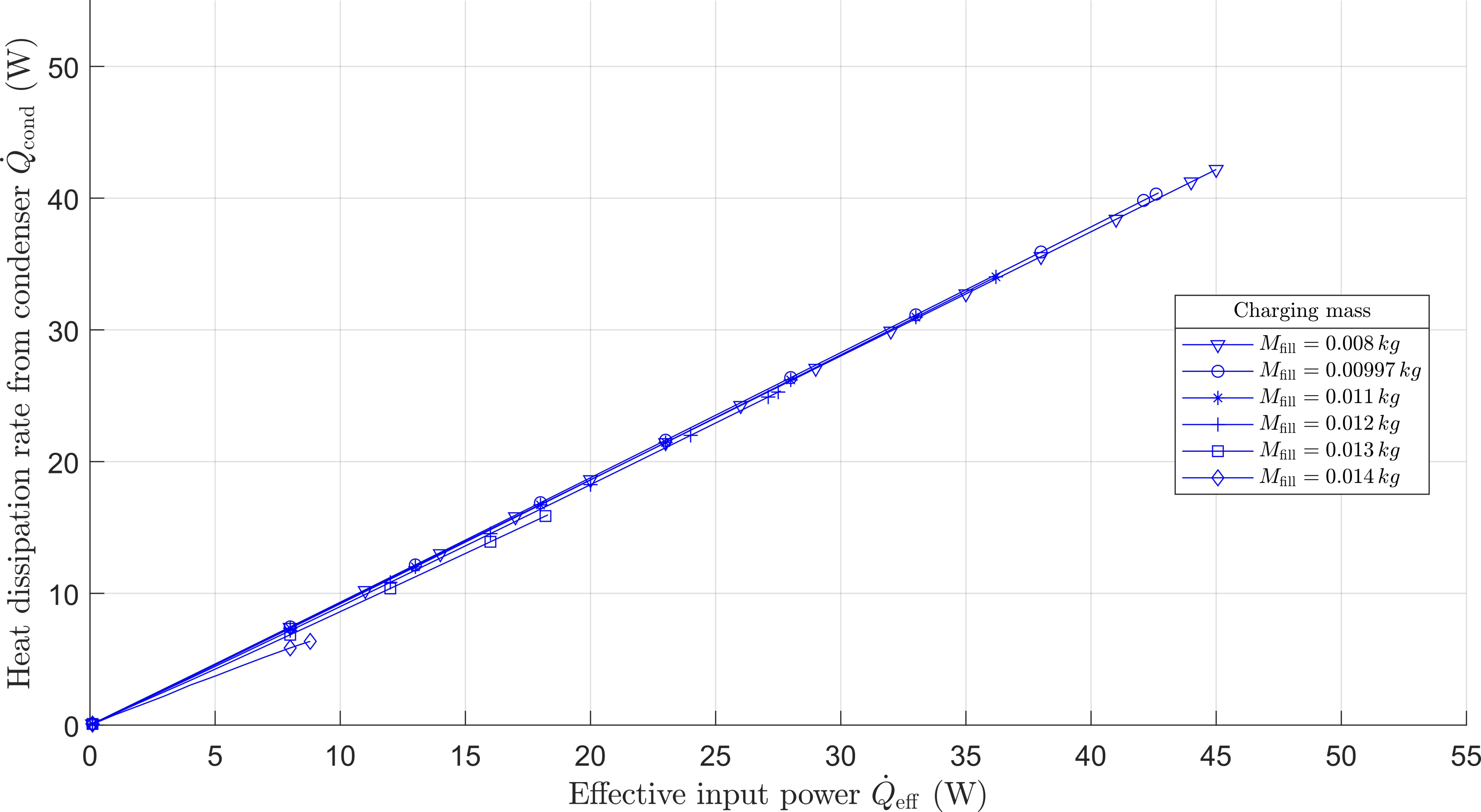}
         \caption{Heat power dissipated from condenser.}
         \label{fig06e}
     \end{subfigure}
      \hspace{0.5mm}  
        \caption{Detailed numerical results for current setup with various charging masses.}
        \label{fig06}
\end{figure}

For our next discussion of these performance curves, we will relate them to the corresponding mass flow rate, two-phase flow section length, condenser pressure drop, wick temperature difference and heat dissipation rate from the condenser,
which are given in Fig. \ref{fig06}.
Ultimately, the simulated performance of the LHP in Fig. \ref{fig05} is the result of the coupled thermal-hydraulic interactions between each of its components.
However, first we will summarize our main observations from Fig. \ref{fig06}. 

Fig. \ref{fig06a} shows the variation of the coolant mass flow rate $\dot{m}$ with the effective input power $\dot{Q}_{\text{eff}}$.
It can be observed that the mass flow rate $\dot{m}$ increases quasi-linearly with the effective input power $\dot{Q}_{\text{eff}}$. 
In addition, a lower charging mass $M_{\text{fill}}$ gives rise to a higher mass flow rate $\dot{m}$, for any given $\dot{Q}_{\text{eff}}$.

Fig. \ref{fig06b} visualizes the variation of the two-phase flow section length $L_{\text{tp}}$ with the effective input power $\dot{Q}_{\text{eff}}$.
We recognize that $L_{\text{tp}}$ remains roughly constant (- it decreases just moderately -) with increasing $\dot{Q}_{\text{eff}}$,
except when the charging mass $M_{\text{fill}}$ becomes rather low.
In that case, we observe first a sharp increase of $L_{\text{tp}}$ with $\dot{Q}_{\text{eff}}$ in the lower range of $\dot{Q}_{\text{eff}}$.
For instance, if {$M_{\text{fill}}=0.008$ kg} or {$M_{\text{fill}}=0.00997$ kg}, this sharp increase occurs when {$\dot{Q}_{\text{eff}} <$ 1 W} (see the curves indicated by the markers $\triangledown$ and $\circ$).
After this sharp increase, $L_{\text{tp}}$ can be considered nearly constant, as $L_{\text{tp}}$ decreases very slowly upon a further increase of $\dot{Q}_{\text{eff}}$.

In Fig. \ref{fig06c}, the dependence of the pressure drop over the condenser $\Delta P_{\text{cond}} = P_{\text{cond,inlet}} - P_{\text{cond,outlet}}$ on the effective input power $\dot{Q}_{\text{eff}}$ is shown. 
The depicted curves share a similar trend: $\Delta P_{\text{cond}}$ first increases quickly with $\dot{Q}_{\text{eff}}$ in the lower range of $\dot{Q}_{\text{eff}}$, 
and then gradually decreases in the higher range of $\dot{Q}_{\text{eff}}$.
At the same time, for a fixed input power $\dot{Q}_{\text{eff}}$, the condenser pressure drop $\Delta P_{\text{cond}}$ increases when the charging mass $M_{\text{fill}}$ is reduced.

In Fig. \ref{fig06d} the relation between the temperature difference across the wick $\Delta T_{\text{wick}} = T_{\text{op}} - T_{\text{cc}}$ and $\dot{Q}_{\text{eff}}$ is illustrated. 
Generally speaking, $\Delta T_{\text{wick}}$ can be seen to increase with $\dot{Q}_{\text{eff}}$, until a maximum value is reached, above which the LHP cannot operate as the corresponding $T_{\text{op}}$ exceeds the preset upper bound as defined by the application temperature limit, e.g. the maximum temperature that is allowed at the die of the electronics component.
When $M_{\text{fill}}=0.011$kg, this maximum value is reached at {24.7 \celsius} (see the curved indicated by the markers $*$).
Yet, a very different behaviour is observed at very low charging masses: for the lowest charging mass of $M_{\text{fill}}=0.008$ kg, $\Delta T_{\text{wick}}$ starts to decrease even after reaching its maximum value of {0.5 \celsius}.

Lastly, in Fig. \ref{fig06e} the relation between $\dot{Q}_{\text{eff}}$ and the heat dissipation rate from the condenser $\dot Q_{\text{cond}}$ is presented.
Overall, the curves show a quasi-linearly increase of $\dot Q_{\text{cond}}$ with $\dot{Q}_{\text{eff}}$.
In addition, a curve for a lower charging mass $M_{\text{fill}}$ always has a larger slope, indicating a larger ratio of $\dot Q_{\text{cond}} / \dot{Q}_{\text{eff}}$. 
A special case is the curve for {$M_{\text{fill}}=0.008$ kg}, whose slope lies between the slopes of the curves for {$M_{\text{fill}}=0.012$ kg} and {$M_{\text{fill}}=0.011$ kg}.

From figures \ref{fig05} and \ref{fig06} we can draw the following conclusions with regard to the LHP's operation.
First of all, we conclude that the condenser is the LHP's main component for dissipating heat into the ambient,

\noindent conform to the LHP's working principle.
Therefore, the LHP's performance is most strongly influenced by the thermal resistance of the coolant-condenser-ambient thermal network.
Even though the evaporator with its integrated compensation chamber serves as an additional component for heat dissipation, 
its thermal resistance with respect to the ambient is too high to allow for significant heat losses.
This appears in particular from Fig. \ref{fig06e}, which shows that the ratio $\dot{Q}_{\text{cond}} / \dot{Q}_{\text{eff}}$ is everywhere higher than 72.1\%, as can be derived from the slope of the plotted curves.
Indeed, the highest contribution of heat dissipated from the evaporator, $\dot{Q}_{\text{cc}} / \dot{Q}_{\text{eff}}$, corresponds to 27.9\% for {$M_{\text{fill}}$ = 0.014 kg}.
Nevertheless, when {$M_{\text{fill}}$ = 0.00997 kg} this contribution is as low as 5.4\%.

Secondly, we conclude that the
main heat dissipation takes place in the two-phase region of the condenser, hence by means of latent heat transfer.
To support this conclusion, we refer to Fig. \ref{fig06a} and Fig. \ref{fig06e}, which confirm that $\dot{Q}_{\text{cond}}$ consists for at least {83\%} of latent heat power $\dot{m} H_{lv}$, 
as $H_{lv} \simeq 2308.4$ kJ/kg within an accuracy of $6\%$ for $T_{\text{op}} \in (T_{\text{amb}}$, 130 \celsius$)$.
The fact that 
$\dot{Q}_{\text{eff}} \simeq \dot{Q}_{\text{cond}}\simeq \dot{m} H_{lv}$ also explains the linear relationship between $\dot{Q}_{\text{eff}}$ and $\dot{m}$
observed in Fig. \ref{fig06a}, since $H_{lv}$ is virtually constant over the entire range of operating conditions.
From a physical point of view, this linear increase of the mass flow rate is caused by an increased generation of vapor per unit of time, proportional to the supplied power $\dot{Q}_{\text{eff}}$.
In turn, this increase in mass flow rate explains the observed initial increase of the condenser pressure drop in Fig. \ref{fig06c}.
The subsequent decrease of $\Delta P _{\text{cond}}$ is due to an abrupt change of the coolant's density and viscosity at higher operating temperatures, which brings us to our next conclusions. 

Our third conclusion is that the 
temperature difference $T_{\text{op}}- T_{\text{amb}}$ essentially drives the heat dissipation towards the ambient.
The reason is that the main heat dissipation takes place in the two-phase region of the condenser, where the coolant temperature is close to $T_{\text{op}}$.
As such, $T_{\text{op}} - T_{\text{amb}}$ is practically the temperature difference between the inside and outside of the condenser.
To support this conclusion, we refer to Fig. \ref{fig06c}, from which we learn that the maximum value of $\Delta P_{\text{cond}}$ is less than {150 Pa} when $M_{\text{fill}} \in (0.009, 0.014)$ kg.
This means that the local pressure in the condenser deviates no more than $5\%$ from $P_{\text{sat}}\left(T_{\text{op}}\right)$ for all operating points considered.
As a result, also the temperature in the condenser will be close to the saturation temperature $T_{\text{op}}$, according to the \textit{Clausius-Clapeyron} equation for the liquid-vapor equilibrium state. 
For instance, when $T_{\text{op}}$ becomes equal to $T_{\text{amb}}$, the vapor temperature in the condenser deviates less than {0.8 \celsius} from $T_{\text{op}}$.
Furthermore, when $T_{\text{op}} =$130 \celsius, the deviation is as small as {0.0186 \celsius}.

From the previous three conclusions, it follows that a higher input power $\dot{Q}_{\text{eff}}$ must come not only with a higher mass flow rate $\dot{m}$, as observed in 
Fig. \ref{fig06a}, but also a higher operating temperature, as observed in Fig. \ref{fig05}.
The higher operating temperature follows from the fact that the driving temperature difference $T_{\text{op}}- T_{\text{amb}}$ must increase when more heat has to be dissipated in the condenser, since $T_{\text{amb}}$ is fixed.
In addition, as mentioned before, also the thermal resistance between coolant and ambient at the condenser plays a crucial role.
However, this resistance is constant to a first approximation, so that the performance curves of $\dot{Q}_{\text{eff}}$ versus $T_{\text{op}}$ in Fig. \ref{fig05} are almost linear, and the LHP is said to operate mainly in \textit{fixed conductance mode}  (FCM) \cite{maydanik2005loop,launay2008analytical}.

Actually, a deeper analysis reveals that the thermal resistance of the coolant-condenser-ambient thermal network does decrease slightly when $T_{\text{op}}$ increases.
The reason is that this resistance is mainly determined by the inverse heat transfer coefficient $1/h_{\text{out}}$ at the outside of the condenser.
The latter decreases as the condenser's outer temperature increases together with $T_{\text{op}}$, according to  \textit{Morgan}'s correlation \cite{morgan1975overall}.
In addition, also the resistance linked to thermal radiation rate in equation (\ref{eq: condenser radiation}) will decrease when the condenser's outer temperature increases together with $T_{\text{op}}$.
As a matter of fact, the condenser's outer temperature will be very close to $T_{\text{op}}$, due to the high heat transfer coefficient $h_{\text{in}}$ inside the two-phase flow region and the relatively thin condenser wall. 
Nevertheless, these combined non-linear effects are small and only impact the thermal resistance per unit length.
As such, the total thermal resistance of the condenser will be barely affected as long as the effective heat transfer length, thus primarily the two-phase section length $L_{tp}$, remains roughly constant.
This is the case for most operating conditions in Fig. \ref{fig06b}. 

Yet, an exception occurs when both the effective input power $\dot{Q}_{\text{eff}}$ and charging mass $M_{\text{fill}}$ are low.
Under such operating conditions, the two-phase section length will vary significantly with $\dot{Q}_{\text{eff}}$, and result in a non-linear operating curve or so-called \textit{variable conductance mode} (VCM) \cite{maydanik2005loop,launay2008analytical}.
In figures \ref{fig05} and  \ref{fig06}, VCM corresponds to the operating range {$\dot{Q}_{\text{eff}} <$ {1 W}} if 
{$M_{\text{fill}} = 0.008$} kg and {$M_{\text{fill}} = 0.00997$ kg}.

In VCM, the mass flow rate is so low, that a small driving temperature difference $T_{\text{op}} -T_{\text{amb}}$ and a relatively short condensation length $L_{\text{tp}}$ are sufficient to cool the vapor coolant back to liquid.
The two-phase section in the condenser will expand upon an increase of $\dot{Q}_{\text{eff}}$, as more heat needs to be dissipated towards the ambient.
This enlargement of $L_{\text{tp}}$ is possible, because VCM occurs when the compensation chamber contains vapor, which can be compressed if subcooled liquid coolant gets displaced into the compensation chamber. 
On the contrary, in FCM, this enlargement is no longer possible, either because the subcooled liquid coolant has already completely filled the compensation chamber (i.e. $\alpha_{\text{cc}}$ = 0), or because the two-phase region has reached the point where it covers the entire condenser (i.e. $L_{\text{tp}} = L_{\text{cond}}$).

There are two conditions that enable the presence of vapor in the compensation chamber during VCM (i.e. $\alpha_{\text{cc}} > 0$).
The first condition is that $T_{\text{cc}}$ becomes only slightly smaller than $T_{\text{op}}$.
Of course, we always have $T_{\text{op}} > T_{\text{cc}}$ during normal operation, as there always occurs a temperature difference over the wick due to its finite thermal conductivity $k_{\text{eff}}$.
Nevertheless, during VCM, $T_{\text{op}} - T_{\text{cc}}$ is particularly low, as we can understand from the energy balance equation (\ref{eq: energy balance cc}):
due to the small input power $\dot{Q}_{\text{eff}}$ and high thermal resistance of the compensation chamber casing, not only the heat loss $\dot{Q}_{\text{cc}}$ from the compensation chamber is small, 
but also the enthalpy difference $(H_{\text{wick/cc}} - H_{\text{cond,outlet}})$ in equation (\ref{eq: energy balance cc}).
This means that heat flux from the wick’s hot side through the wick is small and thus also the temperature difference over the wick $T_{\text{op}} -T_{\text{cc}}$ is minor, as visible in Fig. \ref{fig06d}.
The second condition is that the pressure $P_{\text{cc}}$ in the compensation chamber becomes close to $P_{\text{op}}$, 
because the pressure along the loop remains nearly uniform when the mass flow rate is so low.
In that case, $T_{\text{sat}}(P_{\text{cc}})$ becomes close to $T_{\text{op}}$.
As both conditions together imply that $T_{\text{cc}} = T_{\text{sat}}(P_{\text{cc}})$ can be achieved, the coolant in the compensation chamber can attain a saturated status in which vapor exists.

The range of of heat loads $\dot{Q}_{\text{eff}}$ over which the former conditions arise, and a vapor state can be maintained in the compensation chamber, is primarily determined by the amount of charging mass $M_{\text{fill}}$.
When the charging mass is too high, the liquid coolant will be too densely compressed to allow for the formation of vapor regions after the condenser outlet.
As such, a higher $M_{\text{fill}}$ will limit the attainable length of the two-phase section $L_{tp}$ already at lower heat loads $\dot{Q}_{\text{eff}}$, and thus increase the LHP's overall thermal resistance. 
Consequently, an increasing $M_{\text{fill}}$ not only shifts the transition point between VCM and FCM to the left in figures \ref{fig05} and  \ref{fig06}, but also shifts the (slope of the) depicted performances curves upwards.
As a remark, we add that when the charging mass is extremely low, 
the transition from VCM to FCM will happen through such a strong enlargement of $L_{\text{tp}}$ over a short range of increasing $\dot{Q}_{\text{eff}}$ that the maximum two-phase length 
$L_{\text{tp}} = L_{\text{cond}}$ is reached.
In that special situation, the vapor in the compensation chamber will have been replaced by saturated liquid instead of subcooled liquid, in contrast to the most common type of FCM.
The former special case of FCM is observable in Fig. \ref{fig05} and in Fig. \ref{fig06b} for {$M_{\text{fill}}$ = 0.008 kg}
when $\dot{Q}_{\text{eff}}$ exceeds {1 W}.

We note that in our current setup, the transition from VCM to FCM occurs at much smaller input powers than reported in the literature (\cite{maydanik2020investigation,chernysheva2014copper,chen2012experimental,chernysheva2015effect}). 
This is mainly attributed to the fact that the current LHP relies on natural convection instead of forced convection to cool the condenser.
The heat transfer coefficient at the outside of the condenser is thus much lower and more dependent on the condenser's surface temperature: even for {$T_{\text{op}}$ = 130 \celsius}, $h_{\text{out}}$ is still lower than {19 $\mathrm{W/m\cdot K}$}.
Consequently, the condenser's cooling capacity per unit length is significantly smaller for a given temperature difference $(T_{\text{op}} - T_{\text{amb}})$, as is the two-phase section length $L_{\text{tp}}$.
This restricts the interval of $\dot{Q}_{\text{eff}}$-values over which VCM operation can be maintained.
 
A second reason for the narrow range of VCM operation is
the rather low thermal conductivity of the wick and casing in the current LHP.
VCM requires a small temperature difference over the wick $\Delta T_{\text{wick}}$, even at higher input powers $\dot{Q}_{\text{eff}}$ (thus higher $\dot{Q}_{\text{cc}}$).
Otherwise, the coolant in the compensation chamber will switch sooner towards a subcooled status, and hence FCM, in order to satisfy the 
energy balance equation (\ref{eq: energy balance cc}).
On the contrary, a higher wick conductivity will postpone the divergence between $T_{\text{cc}}$ and $T_{sat}(P_{\text{cc}})$ towards higher input powers and therefore extend the VCM range.

The results in this section have shown the influence of the charging mass on the system performance and the components. The related mechanisms have been analysed, which are interrelated among coolant status, operating mode (VCM and FCM), and cooling intensity. These are closely linked to the operational limits and operational envelopes, as will be shown in next subsection.

\subsubsection{System operational limits and the operational envelope}
\label{subsubsec: operational envelope}
The simulated operational limits for the LHP in our setup are plotted in orange on top of the performance curves in Fig. \ref{fig05}.
Of the nine limits listed in section \ref{subsec: limits}, only the application temperature limit $T_{\text{app}}^{UL}$, the liquid-filling limit $\dot{Q}_{\text{fill}}^{LL}$,  
the heat dissipation limit $\dot{Q}_{\text{disp}}^{UL}$ and the heat leakage limit $\dot{Q}_{\text{hl}}^{UL}$ restrict the LHP's operating range for all possible $M_{\text{fill}}$.
These four limits form a triangle-like operational envelope in Fig. \ref{fig05}.
We remark that the entrainment limit and sonic limit are not shown, as they fall far outside the upper range of allowable input powers $\dot{Q}_{\text{eff}}$.
Further, the viscous limit and the coolant limit are not active.

The pre-set application temperature {$T_{\text{app}}^{UL}$ = 130 \celsius} forms the top of the triangle-like operational envelope in Fig. \ref{fig05}.
The left boundary of the operational envelope is formed by the liquid-filling limit $\dot{Q}_{\text{fill}}^{LL}$. 
It is the collection of the lowest possible input powers $\dot{Q}_{\text{eff}}$ for all specified operating temperatures $T_{\text{op}}$.
As we mentioned in section \ref{subsec: limits}, this limit occurs when $M_{\text{fill}}$ is so high that the entire LHP system is filled with liquid. 
Since the liquid-vapor interface in the wick has disappeared, no capillary force is generated to drive the coolant circulation.  
Therefore, there is almost no heat dissipation from the condenser: $\dot{Q}_{\text{cond}} \simeq 0$.
All input power is dissipated to the environment by means of heat conduction, mainly through the wick’s hot side into the coolant and via the casing's outer surface: $\dot{Q}_{\text{eff}} \simeq \dot{Q}_{\text{cc}}$.
The corresponding conductive thermal resistance is very high, which explains the steepness of the $\dot{Q}_{\text{fill}}^{\text{LL}}$-line. 

The lower right boundary of the envelope, which represents the upper bound of input powers $\dot{Q}_{\text{eff}}$ for each specified $T_{\text{op}}$, is formed by 
the heat-leakage limit $\dot{Q}_{\text{hl}}^{UL}$ when $\dot{Q}_{\text{eff}}$ is below {1 W}.
This boundary part corresponds to VCM operation at the lowest filling rates {$M_{\text{fill}}$ = 0.008 kg} and {$M_{\text{fill}}$ = 0.00997 kg}.
The explanation is that this limit of nearly superheated coolant in the compensation chamber can only become active if the coolant in the compensation chamber is saturated vapor.
Above the limit, superheated coolant must form because of the associated increase in the mass flow rate $\dot{m}$ upon an increase in $\dot{Q}_{\text{eff}}$ at fixed $T_{\text{op}}$. 
This leads to an increased pressure drop $\Delta P_{\text{cond}}$ over the condenser, which in turn reduces
the pressure $P_{\text{cc}}$ and local saturation temperature $T_{\text{sat}}(P_{\text{cc}})$ in the compensation chamber.
As such, superheating of the coolant will occur due to this reduction in $T_{\text{sat}}(P_{\text{cc}})$, causing a stop of coolant circulation.

For $\dot{Q}_{\text{eff}}$ above {1 W}, the right boundary of the operational envelope is determined by both the heat-leakage limit $\dot{Q}_{\text{hl}}^{UL}$  
and the heat-dissipation limit $\dot{Q}_{\text{disp}}^{UL}$.
This boundary part corresponds to FCM operation at the lowest filling rate {$M_{\text{fill}}$ = 0.008 kg}, since  
the condensation flow occupies the entire condenser ($L_{\text{tp}} = L_{\text{cond}}$) when this limit of full condenser utilization
is active.
Also the evaporator is fully utilized, because $\dot{Q}_{\text{cc}}$ can no longer increase once the coolant temperature in the compensation chamber has reached its maximum
value $T_{\text{sat}}(P_{\text{cc}})$.

We remark that in principle, also
the liquid-filling limit $\dot{Q}_{\text{fill}}^{UL}$
could be activated for low charging masses, although it is not the case here. It will effect the right boundary of the envelope. 
Activation is defined when $\alpha_{\text{cc}} \geq 0.5$, which occurs at a  saturated coolant temperature. This situation inhibits the coolant circulation in the system at the low effective power range. This low power range will become larger the lower $M_{\text{fill}}$.

The key takeaway from the operational envelope in Fig. \ref{fig05} is that 
the LHP's performance for given operating conditions is mainly decided by the charging mass $M_{\text{fill}}$.
It is beneficial to reduce $M_{\text{fill}}$ as much as possible within the limits imposed by heat leakage, heat dissipation, and possibly liquid-filling.
After all, this way we achieve a lower operating temperature $T_{\text{op}}$ for a given input power $\dot{Q}_{\text{eff}}$.
In particular, this is because a lower charging mass $M_{\text{fill}}$ allows for a longer condensation section $L_{\text{tp}}$ and thus a lower overall thermal resistance. 

The heat-leakage limit is the primary constraint when the LHP is operated in VCM.
Therefore, the LHP's performance in terms of obtaining a higher $\dot{Q}_{\text{eff}}$ for a given $T_{\text{op}}$ during VCM can be improved by keeping the coolant in the compensation chamber subcooled. 
This requires to either increase the saturation temperature, or reduce the coolant temperature there.
The first option typically has little effect, as $\Delta P_{\text{cond}}$ is small anyway.
The last option can be achieved by employing a wick (and casing) with a lower thermal conductivity, or improving the heat dissipation rate from the evaporator and compensation chamber casing.

The heat-dissipation limit is the primary constraint when the LHP is operated in FCM. 
This means that the LHP's performance during FCM can be improved by increasing the cooling intensity outside the condenser, for instance, by attaching fins to the condenser surface,
or by relying on forced convection. In that case, a detailed parameter investigation is necessary to enlarge the operational envelope for improving the design.

\section{Conclusion}
\label{sec: conclusion}
In this paper, an LHP with a squared wick and evaporator is established with the aim to cool electronic components. The setup operates successfully with a natural convection cooled condenser and proves that an LHP is also in this application a promising device. A physics-based 1D numerical model for LHPs is elaborated and successfully validated by the test results. The two-phase flow section length measurements in the condenser for five different heat loading conditions serve for the calibration of the condenser heat transfer correlations. An overall energy balance analysis proves the necessity of the tuning parameters in heat transfer coefficients, for both the two-phase flow inside the condenser and the natural convection outside the condenser. After correcting these heat transfer coefficients and after including the volume of non-condensable gases, the physics-based 1D numerical model for LHPs can reproduce the test measurements within experimental accuracy.

The performance curves for the current setup for different charging masses are provided by the 1D model. Given the low thermal conductive materials for the wick and the evaporator wall, the temperature of the coolant in the compensation chamber is mostly lower than the local saturation temperature for the majority of the input powers. In combination with the low pressure drop in the condenser, the fixed conductance mode is dominant for this setup. The variable conductance mode does occur only when the input power is low ({$<$ 1 W}) and charging mass is in the lower range ({$<$ 0.01 kg}).
For higher input power and with this lower charging mass range both the CC and the condenser can be fully utilized used for heat dissipation, while the system is still operating in the fixed conductance mode.

Finally, the operational envelope for the current setup is assessed for all possible charging masses. It is unveiled that for a certain operating temperature, the constraint on the maximum input power that can be dissipated by this setup is mainly caused by the heat dissipation limit, which then induces the heat-leakage limit. While when the input power is small (lower than {1 W}) and charging mass is low, it is the heat-leakage limit being the constraint. The liquid-filling limit might become active depending on the charging mass, but does not occur for the studied charging mass. The best power and thermal resistance performance is achieved for the lowest charging masses. However, the lower value for the charging mass is limited by the heat dissipation and heat-leakage limit. For the studied test setup the charging mass brings the best performance for charging mass around 0.01 kg. On the other hand, to extend the operational envelope, the most efficient way is to increase the cooling intensities outside the condenser.

Although this physics-based 1D LHP model well predicts the behavior of the present setup, several steps are left for further research. For the experimental part, a dedicated method to evaluate the effective input power and the amount of NCG could further reduce the model validation uncertainties.
For the modelling part, the influence of the evaporator wall should be included, especially for LHPs with highly thermal conductive material in real applications. Finally, a 3D detailed study of the wick structure, its shape and its consequences for the overall system performance might further reveal improvements to be made to the set-up. 

\appendix
\setcounter{table}{0}
\section{Thermal resistance for the heat losses from the compensation chamber}
\label{app: thermal resistance from coolant in cc to ambient}
The thermal resistance associated with the heat losses from the cuboid compensation chamber to the ambient consists of three terms:
\begin{equation}
\label{eq: Rcc total}
\frac{1}{R_{\text{cc}}} = 
\frac{1}{R_{\text{cc,1}}} + 
\frac{1}{R_{\text{cc,2}}} + 
\frac{1}{R_{\text{cc,3}}}
\end{equation}
The first term stems from the thermal path through the longest side walls of the compensation chamber.
Along this path $1$, there is heat conduction through  the casing 
and insulation material, next to passive heat convection at the outside of the insulation, so 
\begin{equation}
R_{\text{cc,1}} = \frac{1}{2} 
\left[
\frac{t_{\text{case,1}}}{k_{\text{case}} A_{\text{case,1}}} +  
\frac{t_{\text{insu}}}{k_{\text{insu}} A_{\text{insu,1}}} +  \frac{1}{h_{\text{insu,1,out}}A_{\text{insu,1,out}} }   
\right] 
\end{equation}
Here, $k_{\text{case}}$ and $k_{\text{insu}}$ denote the thermal conductivities of the polycarbonate casing and insulation material.
$A_{\text{case,1}}$, $A_{\text{insu,1}} $, $A_{\text{insu,1,out}} $ denote their respective (inner and outer) 
side areas, while $t_{\text{case,1}}$
and $t_{\text{insu}}$  denote their thicknesses.
$h_{\text{insu,1,out}}$ is the passive heat convection coefficient.
The factor ${\rm 1/2}$ appears in the right-hand side because two parallel side walls should be considered. 

Similarly, the second term $R_{\text{cc,2}}$ is linked to the
thermal path $2$ formed by the shortest side walls:
\begin{equation}
R_{\text{cc,2}} = \frac{1}{2} 
\left[
\frac{t_{\text{case,2}}}{ k_{\text{case}} A_{\text{case,2}}} +  \frac{t_{\text{insu}}}{ k_{\text{insu}} A_{\text{insu,2}}} +  \frac{1}{h_{\text{insu,2,out}}A_{\text{insu,2,out}} }   
\right] 
\end{equation}

The last term is due to the thermal path $3$ through the top wall:
\begin{equation}
R_{\text{cc,3}} = 
\frac{t_{\text{case,3}}}{ k_{\text{case}} A_{\text{case,3}}} +  \frac{t_{\text{steel}}}{ k_{\text{steel}} A_{\text{steel,3}}} + \frac{t_{\text{insu}}}{ k_{\text{insu}} A_{\text{insu,3}}} +  \frac{1}{h_{\text{insu,3,out}} A_{\text{insu,3,out}} } 
\end{equation}
In this last resistance, the presence of the 
the steel plate between the Polycarbonate casing and insulation material is taken into account, assuming its surface area $A_{\text{steel}}$ the same as the outer surface of the casing's top wall.
This steel plate is used to connect the heating simulator and the LHP (indicated by (\textit{path} 3, \textit{layer} 2) in Table \ref{tab_app_01}).

To evaluate $h_{\text{insu,1,out}}$, $h_{\text{insu,2,out}}$ for the vertical walls, the average \textit{Nusselt} number is evaluated by integrating the \textit{Squire-Eckert} equation over the height of the vertical walls \cite{lienhard2005heat}
\begin{equation}\label{vertical passive heat transfer coefficient}
\overline{Nu}_{\text{V}} = 
0.678 \cdot Ra^{0.25} \big(\frac{Pr}{0.953 + Pr} \big) ^{0.25} 
\end{equation}
The characteristic length used to evaluate the \textit{Rayleigh} number $Ra$ and \textit{Nusselt} number $\overline{Nu}_{\text{V}}$ is the height of the insulation plate.

To evaluate $h_{\text{insu,3,out}}$ for the horizontal top walls, the surface is assumed to be isothermal and the \textit{Nusselt} number is given by \cite{goswami2004crc}
\begin{equation} \label{horizontal passive heat transfer coefficient}
\overline{Nu}_{\text{H}} = 
\frac{0.560 \cdot Ra^{0.25}}{[1 + (0.492/Pr)^{9/16}]^{4/9}} 
\end{equation} 
The characteristic length used to evaluate the \textit{Rayleigh} number $Ra$ and \textit{Nusselt} number $\overline{Nu}_{\text{H}}$ is the ratio of the horizontal surface's area to its perimeter.

For the geometrical details and  materials properties, we refer to  Table \ref{tab_app_01}.

{\fontsize{7}{10}\selectfont
\setlength\tabcolsep{3.0pt}
\centering
\begin{longtable}[H]{p{1.5cm}p{2.35cm}p{1.8cm}p{0.05cm}p{1.5cm}p{2.35cm}p{1.8cm}}

\caption{The main parameters of the experimental setup components.}\label{tab_app_01}
\\ \cline{1-3} \cline{5-7}
\multirow{2}{1.5cm}{Wick}                                       &  \multirow{2}{2.35cm}{Width/Length/ Height}                                                       &   \multirow{2}{1.8cm}{10.0/10.0/4.5 mm}      
&&   \multirow{6}{2.0cm}{PC plate}                  &  Thickness                                                            & 8.0 mm    
\\
        &           &           &&          &    \multirow{2}{2.35cm}{Thermal conductivity}                                                 & \multirow{2}{2.35cm}{0.19 ${\rm W/m\cdot K}$}   
\\ \cline{1-3} 

\multirow{3}{1.5cm}{Heating simulator}   &  Active surface area       &   1 ${\rm cm^2}$      
&&              &           &
\\ 
        &  \multirow{2}{2.35cm}{Side wall Width/ Height}         &   \multirow{2}{1.8cm}{20.0/20.0 mm}        &&          &   \multirow{2}{2.35cm}{Specific heat capacity}        &   \multirow{2}{2.35cm}{1090 ${\rm J/kg\cdot K}$}
\\ 
        &           &           &&          &           &
\\ 
        & Total volume          &  8.5 ${\rm cm^3}$         &&          &      Density     &   1220 ${\rm kg/m^3}$
\\ \cline{1-3} \cline{5-7}

\multirow{3}{1.6cm}{Fixture side wall (\textit{path} 1)}                                       & Material                                                       &   PEEK, Teflon      
&&    \multirow{6}{1.6cm}{Ceramic fibre plate}          &    Thickness       &  10 mm
\\ 
        &  \multirow{2}{2.35cm}{Length/Height/ Thickness}         &   \multirow{2}{1.8cm}{56.0/40.0/3.0 mm}        &&          &    \multirow{2}{2.35cm}{Thermal conductivity}    & \multirow{2}{2.35cm}{0.085 ${\rm W/m\cdot K}$}   
\\ 
        &           &           &&          &           &
\\ \cline{1-3} 

\multirow{3}{1.6cm}{Fixture side wall (\textit{path} 2)}                                       & Material                                                       &   PEEK, Teflon      
&&              &    \multirow{2}{2.35cm}{Specific heat capacity}        &   \multirow{2}{2.35cm}{1130 ${\rm J/kg\cdot K}$}
\\  
        &  \multirow{2}{2.35cm}{Length/Height/ Thickness}         &   \multirow{2}{1.8cm}{34.0/40.0/11.0 mm}        &&          &           &
\\ 
        &           &           &&          &     Density      &   300 ${\rm kg/m^3}$
\\ \cline{1-3} \cline{5-7}

\multirow{6}{1.6cm}{Fixture bottom wall (\textit{path} 3)}                                       & \textit{layer} 1 material                                                       &   PEEK, Teflon      
&&   \multirow{6}{1.6cm}{Polymer foil}          &    Thickness       &  3 mm
\\ 
        &  \multirow{2}{2.35cm}{Length/Height/ Thickness}         &   \multirow{2}{1.8cm}{56.5/40.0/15.0 mm}        &&          &    \multirow{2}{2.35cm}{Thermal resistance}    & \multirow{2}{2.35cm}{0.083 ${\rm m^2\cdot K/W}$}  
\\ 
        &           &           &&          &           &
        
\\ 
                & \textit{layer} 2 material                                                       &   Steel      
&&              &       \multirow{2}{2.35cm}{Specific heat capacity}        &   \multirow{2}{2.35cm}{1300 ${\rm J/kg\cdot K}$}
\\ 
        &  \multirow{2}{2.35cm}{Length/Height/ Thickness}         &   \multirow{2}{1.8cm}{56.0/40.0/2.5 mm}        &&          &           &
\\ 
        &           &           &&          &   Density        &  25 ${\rm kg/m^3}$
\\ \cline{1-3} \cline{5-7}

\multirow{6}{1.6cm}{Evaporator bottom plate}        &   \multirow{2}{2.1cm}{Material}        &  \multirow{2}{2.1cm}{Oxygen-free copper}         &&     \multirow{6}{1.6cm}{Polymer plate}     &      Thickness       &    12.0 mm
\\ 
        &           &           &&              &     \multirow{2}{2.35cm}{Thermal conductivity}     &  \multirow{2}{2.35cm}{0.043 ${\rm W/m\cdot K}$}  
\\  
        &  \multirow{2}{2.1cm}{Fins/Grooves number}         &  \multirow{2}{2.1cm}{4/3}         &&          &       & 
\\ 
        &           &           &&          &     \multirow{2}{2.35cm}{Specific heat capacity}      &     \multirow{2}{2.35cm}{1300 ${\rm J/kg\cdot K}$}
\\ 
        &  \multirow{2}{2.35cm}{Fin/Groove Width /Length/Height}                  &  \multirow{2}{2.3cm}{1.05/10.0/5.5 \\mm}            &&          &            &   
\\
        &           &           &&          &      Density         &  25 ${\rm kg/m^3}$
\\ \cline{1-3} \cline{5-7}

\multirow{2}{1.6cm}{Evaporator casing}        & \multirow{2}{2.1cm}{Material}          &  \multirow{2}{2.1cm}{PC plates}         &&     \multirow{5}{1.6cm}{Teflon}      &   \multirow{2}{2.35cm}{Thermal conductivity}     &  \multirow{2}{2.35cm}{0.025 ${\rm W/m\cdot K}$} 
\\ 
        &           &           &&          &           &
\\ \cline{1-3} 

\multirow{2}{1.6cm}{Casing side wall (\textit{path} 1)}        & \multirow{2}{2.1cm}{Length/Height/ Thickness}          & \multirow{2}{2.1cm}{31.5/71.0/8.0 mm}          &&          &        \multirow{2}{2.35cm}{Specific heat capacity}      &     \multirow{2}{2.35cm}{1000 ${\rm J/kg\cdot K}$}
\\ 
        &           &           &&          &           &
\\ \cline{1-3}

\multirow{2}{1.6cm}{Casing side wall (\textit{path} 2)}         & \multirow{2}{2.1cm}{Width/Height/ Thickness}          & \multirow{2}{2.1cm}{11.5/71.0/8.0 mm}          &&          &         Density  &  2200 ${\rm kg/m^3}$
\\ \cline{5-7}
        &           &           &&    \multirow{5}{1.6cm}{PEEK}      &    \multirow{2}{2.35cm}{Thermal conductivity}     &  \multirow{2}{2.35cm}{0.25 ${\rm W/m\cdot K}$} 
\\ \cline{1-3} 

\multirow{6}{1.6cm}{Casing top wall (\textit{path} 3)}        &  \textit{layer} 1 material         &  PC plates         &&          &           &
\\ 
        & \multirow{2}{2.1cm}{Length/Width/ Thickness}          & \multirow{2}{2.1cm}{11.5/15.5/10.0 mm}          &&          &   \multirow{2}{2.35cm}{Specific heat capacity}      &     \multirow{2}{2.35cm}{320 ${\rm J/kg\cdot K}$}
\\ 
        &           &           &&          &           &
\\ 

       &  \textit{layer} 2 material         & Steel          &&          &      Density     &  1320 ${\rm kg/m^3}$
\\ \cline{5-7} 
          & \multirow{2}{2.1cm}{Length/Width/ Thickness}          & \multirow{2}{2.1cm}{56.0/40.0/2.5 mm}          &&    \multirow{5}{1.6cm}{Copper}      &    \multirow{2}{2.35cm}{Thermal conductivity}     &  \multirow{2}{2.35cm}{385 ${\rm W/m\cdot K}$} 
\\ 
        &           &           &&          &           &
\\ \cline{1-3} 

\multirow{5}{1.9cm}{Compensati- on chamber}                       & Volume                                                                & 9.09 ${\rm cm^{3}}$        &&          &       \multirow{2}{2.35cm}{Specific heat capacity}      &     \multirow{2}{2.35cm}{385 ${\rm J/kg\cdot K}$}
\\ 
         & \multirow{2}{2.3cm}{Length/Width/
         Height}   & \multirow{2}{2.1cm}{15.5/11.5/51.0 mm}         &&          &           &
\\  
        &           &           &&          &     Density      &  8960 ${\rm kg/m^3}$
\\ \cline{5-7}  
        & \multirow{2}{2.3cm}{Cuboid formed inside casing}                      & \multirow{2}{2.1cm}{15.5/11.5/61.0 mm}  &&    \multirow{5}{1.6cm}{Steel}      &    \multirow{2}{2.35cm}{Thermal conductivity}     &  \multirow{2}{2.35cm}{45 ${\rm W/m\cdot K}$} 
\\  
        &           &           &&          &           &
\\ \cline{1-3} 

\multirow{4}{1.9cm}{Condenser \\pipe}  &   Material        &  PFA         &&          &          \multirow{2}{2.35cm}{Specific heat capacity}      &     \multirow{2}{2.35cm}{420 ${\rm J/kg\cdot K}$}
\\  
        & \multirow{3}{2.3cm}{Inner diameter/ Outer diameter/ Length}          & \multirow{3}{2.1cm}{2.67/4.00/1010 mm}  &&          &           &
\\  
        &           &           &&        &     Density        &   9050 ${\rm kg/m^3}$
\\  \cline{5-7}
        &           &           
\\ \cline{1-3} 
\multirow{6}{2.0cm}{PFA}      &  \multirow{2}{2.0cm}{Thermal conductivity}    & \multirow{2}{2.0cm}{0.209 ${\rm W/m\cdot K}$} 
\\ 
        &           &           
\\  
        & \multirow{2}{2.0cm}{Specific heat capacity}    & \multirow{2}{2.0cm}{1047 ${\rm J/kg\cdot K}$}     
\\  
        &           &          
\\ 
        &   Density & 2170 ${\rm kg/m^3}$   
\\ 
        &   Emissivity  & 0.97   
\\ \cline{1-3} 

\end{longtable}
}
 
\section{Thermal resistance for the heat losses from the copper heating block}
\label{app: thermal resistance from copper heating block to ambient}
The overall thermal resistance $R_{\text{heat}}$ from the copper heating block to the ambient is similar in nature to $R_{\text{cc}}$
from (\ref{eq: Rcc total}) as there are again three paths for heat losses:  two through the parallel sides and one via the bottom surface.
We have ignored the thermal contact resistance. The conductive resistance of the fixture and the insulation material 
follows from the dimensions enlisted in Table \ref{tab_app_01}.

\section*{CRediT authorship contribution statement}
\textbf{Xu Huang:} Conceptualization, Methodology, Software, Investigation, Data Curation, Writing - Original Draft, Visualization, Funding acquisition.
\textbf{Geert Buckinx:} Methodology, Writing - Original Draft, Writing - Review \& Editing, Funding acquisition.
\textbf{Maria Rosaria Vetrano:} Methodology, Validation, Resources, Writing - Review \& Editing, Supervision.
\textbf{Martine Baelmans:} Conceptualization, Methodology, Resources, Writing - Review \& Editing, Supervision, Project administration.

\section*{Declaration of Competing Interest}
The authors declare that they have no known competing financial interests or personal relationships that could have appeared to influence the work reported in this paper.

\section*{Data availability}
Data will be made available on request.

\section*{Acknowledgement}
The author greatly appreciated the support from the China Scholarship Council under Grant 201506030103. This
work was partly funded by the Research Foundation | Flanders (FWO) through G. Buckinx's post-doctoral project grant 12Y2919N. 

\bibliographystyle{elsarticle-num} 
\bibliography{reference_preprint}
\end{document}